\documentclass[fleqn,10pt]{wlscirep}
\usepackage[utf8]{inputenc}
\usepackage[T1]{fontenc}
\usepackage{xcolor}
\usepackage{float}
\usepackage{hyperref}
\usepackage{lineno}

\definecolor{mypink}{RGB}{220, 38, 127}
\definecolor{myblue}{RGB}{0, 114, 178}
\definecolor{liteblue}{RGB}{86, 180, 233}

\title{Discovering a low-dimensional temperature control architecture across animals}

\author[1,*]{Cody E. FitzGerald}
\author[2]{Andrew J. Engedal}
\author[1,3]{Niall M. Mangan}
\affil[1]{Department of Engineering Sciences and Applied Mathematics, Northwestern University, Evanston, IL 60208, USA}
\affil[2]{Lewis-Sigler Institute of Integrative Genomics, Princeton University, Princeton, NJ 08540, USA}
\affil[3]{NSF-Simons National Institute for Theory and Mathematics in Biology, Chicago, IL  60611, USA}
\affil[*]{cody.fitzgerald@northwestern.edu}


\begin{abstract}
Hibernation is an adaptation to extreme environmental seasonality that has been studied for almost 200 years, but our understanding of the underlying physiological system remains lacking due to the partially observed nature of the system. During hibernation, small mammals, such as the Arctic ground squirrel, exhibit dramatic oscillations in body temperature, typically one of the only physiological states measured, of up to 40 $^{\circ}$C. These spikes are known as interbout arousals and typically occur 10-20 times throughout hibernation. The physiological process that drives interbout arousals is unknown, but two distinct macro-scale mechanisms have been hypothesized. Using model selection for partially observed systems and classical dynamical systems theory, we are able to differentiate between these two hypotheses using only body temperature data recorded from a free-ranging Arctic ground squirrel, and show that our model can capture the broad features of the observed seasonal physiological transitions. We then modify our discovered physiological model of Arctic ground squirrel to include internally-encoded environmental information and find that we can qualitatively match body temperature data recorded from a wide range of species, including a bird, a shrew, and a bear, which also dynamically modulate body temperature. Our results suggest that a low-dimensional, environmentally sensitive core regulator could control body temperature across a diverse range of species---a new understanding of the physiological organization across species. While the findings presented here are applicable to thermophysiology, the general modeling procedure is applicable to time series data collected from partially observed biological, chemical, physical, mechanical, and cosmic systems for which the goal is to elucidate the underlying mechanism or control structure.
\end{abstract}
\begin{document}

\flushbottom
\maketitle
%
%
\thispagestyle{empty}


\section{Introduction}

Scientists have studied hibernation for nearly two hundred years \cite{hall1832hybernation} in the hopes of gaining insight into this complex physiological system and to extend our own physiology. Hibernation research has a wide range of practical applications to human health, including cardiac arrest, organ preservation, and longevity. Inducing hibernation-like states into the range of human physiology is also under study, as this may help realize our species' goals of long-distance human space travel \cite{petit2018hibernation,chouker2019hibernating}. Interest in hibernation and dormancy states has transcended science and also fills our literature. Proust writes of ``...the comfort of reclusion, the poetry of hibernation...'' in \textit{A Swan's Way}. A more recent novel, \textit{My Year of Rest and Relaxation}, follows a New York City woman who entered a months-long hibernation-like state. But for all this sustained scientific and broader interest, our understanding of dynamic temperature regulation (known as the thermophysiological system) that facilitates hibernation remains limited, in part because few physiological states of the system can be observed experimentally. Typically, body temperature is the only physiological measurement that can be recorded from free-ranging organisms, though surely other states of the physiological system must contribute to thermoregulation. In applied mathematics, this is known as a \textit{partially observed system}, and inferring models for this class of system is technically challenging and an active area of research \cite{dahsi,stepaniants2024discovering}. Here, we combine body temperature measurements with automated model discovery and classical dynamical systems theory to develop a mechanistic understanding of small mammalian thermophysiology. We then show that our discovered model acts as a low-dimensional control architecture capable of describing observed temperature dynamics across multiple species. 

\begin{figure}[h!]
    \begin{center}
        \includegraphics[width=0.9\textwidth]{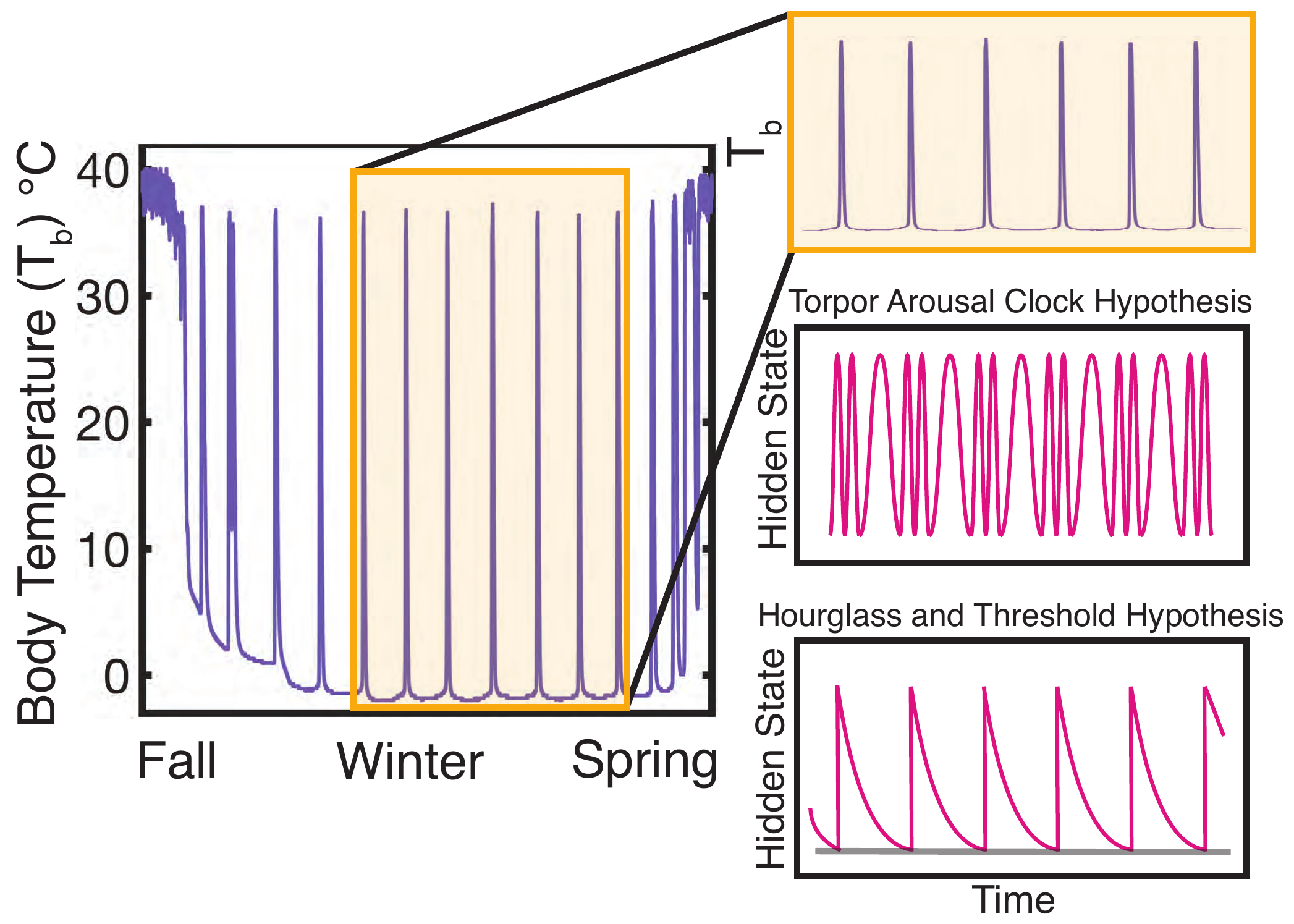} 
    \end{center}
    \caption{The physiological mechanism driving the spiking behavior observed in the body temperature of the hibernating Arctic ground squirrel is unknown. Two distinct mechanistic hypotheses describe the dynamics of the hidden physiological mechanism. The distinguishing feature is whether the oscillation has a variable period (Torpor Arousal Clock Hypothesis) or a fixed period (Hourglass and Threshold Hypothesis, assuming a relatively constant metabolite decay rate and fixed physiological threshold shown in gray). Left: Body temperature recording of the Arctic ground squirrel during hibernation from \cite{chmura}. Right: In the Torpor Arousal Clock Hypothesis, the hidden physiological driver of interbout arousals has a variable period that depends on the body temperature of the organism (see Figure 3 in \cite{malan}). When the body temperature of the organism is high, the hidden physiological state is predicted to have a short period. When the body temperature of the organism is low, the hidden physiological state is predicted to have a long period. In the Hourglass and Threshold Hypothesis, the hidden physiological driver of interbout arousals has a constant period, assuming a constant metabolite decay rate and fixed physiological threshold (shown in gray) \cite{ruf2022}.}
    \label{fig:0}
\end{figure}

We begin our search for a low-dimensional temperature controller by studying the Arctic ground squirrel, a species that has been well-studied and exhibits seasonal physiological changes representative of many small mammalian hibernating species. The Arctic ground squirrel lives in the extremes of the high North American Arctic, and it experiences dramatic seasonal physiological changes. In a typical year, an Arctic ground squirrel remains above ground for approximately four months and hibernates underground for about eight months. During the summer months, the Arctic ground squirrel is homeothermic, meaning its body temperature hovers around 37 $^{\circ}$C with minimal daily fluctuations caused by activity. A seasonal thermoregulatory transition occurs as summer merges into fall, and the Arctic ground squirrel enters hibernation (Figure \ref{fig:0}). At the start of hibernation, the Arctic ground squirrel makes a thermoregulatory transition from homeothermy to the \textit{hibernation thermoregulatory regime} in which body temperature alternates between homeothermy near or below 0 $^{\circ}$C and occasional returns to euthermic conditions near 37 $^{\circ}$C. These returns to euthermic conditions are known in the biological literature as ``interbout arousals.'' During hibernation, the body temperature of the Arctic ground squirrel can be as low as -2.9 $^{\circ}$C. The organism can lower its body temperature below the freezing point through a process called super-cooling \cite{barnes1989freeze}. Toward the end of hibernation, the Arctic ground squirrel makes the reciprocal thermoregulatory switch from the hibernation thermoregulatory regime to homeothermy, and the annual cycle continues. An open challenge is determining what underlying mechanisms control these dramatic changes in body temperature. 

The physiological basis of interbout arousals is unknown. Two hypotheses have been put forward: the Torpor Arousal Clock Hypothesis \cite{malan} and the Hourglass and Threshold Hypothesis \cite{ruf2022} (Figure \ref{fig:0}, right). In the Torpor Arousal Clock Hypothesis, the timing of interbout arousals is controlled by a non-temperature-compensated circadian clock. Critically, this hypothesis suggests that the hidden physiological state will oscillate and the period of the oscillation will vary dynamically as a function of body temperature (see Figure \ref{fig:0}, middle right and Figure 3 in \cite{malan}). In the Hourglass and Threshold Hypothesis, interbout arousals are caused by an unknown metabolite decreasing to a physiological threshold (gray in Figure \ref{fig:0}, bottom right), which then triggers an interbout arousal. The molecule is thought to be replenished only when the organism rewarms during hibernation (see the sharp peaks in body temperature, Figure \ref{fig:0}, blue). If the decay rate of the metabolite or the value of the physiological threshold changes over time, the oscillatory dynamics of the hidden physiological driver may have a variable period \cite{ruf2022}. However, if the decay rate of the metabolite and the physiological threshold are relatively constant, we would expect the dynamics of the hidden physiological driver to oscillate with a relatively constant period (Figure \ref{fig:0}, bottom right). Mathematically, we will be able to differentiate between these two hypotheses if the unknown physiological state that controls body temperature oscillates with a constant period, as in the Hourglass and Threshold Hypothesis, or a variable period, as in the Torpor Arousal Clock hypothesis. To infer the dynamics of the hidden physiological driver using model selection, we need to simplify and modify previous models of small mammalian hibernation.




Relatively few mathematical models of hibernation and thermophysiology exist, and those that do exist tend to be either relatively complex and thus difficult to validate against the existing data or statistical in nature. Hampton and Andrews put forward a mechanistic model of small mammalian hibernation \cite{hampton2007}. Their model puts forward a mechanism that triggers interbout arousals and can be viewed as a version of the Hourglass and Threshold Hypothesis. However, the model formulation is very high-dimensional and cannot be validated using the available experimental data. The authors also developed a complex physiological model to infer metabolic rate from other physiological data \cite{hampton2010circulation}. Others have used statistical frameworks to examine features of hibernation. Oro and Freixas used statistical modeling to show that air temperature drives ``flickering,'' a behavior that occurs as the organism transitions between homeothermy (constant body temperature) and the hibernation thermoregulatory regime (spiking body temperature) \cite{oro2021}. A different study used Generalized Harmonic Analysis to investigate interbout arousals in small hibernators \cite{gibo2024frequency}. We seek to find the right level of mathematical description for hibernation physiology that balances interpretability and preserves the ability to validate the model against the existing data. This is challenging due to natural and experimental limitations intrinsic to the system. 

Hibernation is a multi-scale physiological adaptation that occurs on the year scale. For this reason, a molecular description of the biological processes that underpin hibernation will be exceedingly difficult to find. It is unlikely we will ever have complete and comprehensive data sets of all relevant neurophysiological subsystems involved in hibernation due to the costs of studying hibernators from both a financial and animal welfare perspective. Most of the data collected from hibernation experiments occurs at the physiological scale rather than the molecular scale. We grappled with the right size and type of model to use in describing this system, and ultimately found a model that is roughly as mechanistic the Fitzhugh-Nagamo model \cite{fitzhugh,nagumo}, which describes neuronal spiking dynamics; the NPZ model \cite{franks1986behavior}, which describes the dynamics of nutrients, phytoplankton and zooplankton; or the Brusselator model \cite{lefever1971chemical,prigogine1967dissipative,prigogine1968symmetry} which describes the Belousov-Zhabotinsky chemical reaction \cite{am1964periodic}. The goal of these models, and perhaps all models, is to describe the essence of the underlying process in an interpretable manner without being completely overrun by the complexity of the system or the curse of dimensionality. Balancing model simplicity, which allows for interpretability, with complexity, which allows for greater realism, is a fine line. The NPZ model is a good example of a model that exhibits this balance. It captures a complex, multi-scale ecological system using just three coupled equations, and can be used to probe how changes or perturbations to one state influence the others, which would be challenging to understand both without any model and with too complex a model. Here, we attempt to find a similar modeling balance. We discover a macro-scale model with interpretable parameters that have physiological meaning, that is complex enough to capture the main features of the data,  that is simple enough to be fit to the existing physiological data, and can be used to guide the search for microscale mechanisms that underpin hibernation. Using our discovered model, we provide new insights into hibernation physiology, seasonal physiological transitions, and the possible conserved physiological organization across species. 

To find this model, we use a recent sparse model selection algorithm for partially observed systems and discover a simple, data-driven model of small mammalian hibernation. We then use this model to differentiate between the Torpor Arousal Clock and Hourglass and Threshold Hypotheses. In our analysis, we recover the dynamics of a hidden physiological state, which has a constant period, lending support to the Hourglass and Threshold Hypothesis. We also use our discovered model to explore seasonal thermoregulatory transitions through a bifurcation analysis, which results in a multi-scale understanding of hibernation that links the neurobiological and physiological scales. As a final test of our discovered Arctic ground squirrel model, we used it to explore a broader, interspecific question: Could a core thermophysiological controller be responsible for regulating similar thermoregulatory behaviors across species? 

During the annual cycle, the Arctic ground squirrel switches between two distinct thermoregulatory behaviors: a stable, regulated body temperature and a dynamically changing body temperature marked by temporal oscillation. Many other organisms exhibit similar transitions between the two thermoregulatory modes, but the timescale of the behaviors and timing of transitions between them is different. Long-term body temperature data sets have been collected from these (and other) species \cite{chmura,toien,miner,shrew}, but it is unclear how the patterns, and thus the underlying systems, relate to one another. A satisfactory model would be capable of capturing a broad range of observed body temperature patterns across species and would further our understanding of physiological organization across species. Previous analyses attempted to categorize body temperature recordings across species using a statistical lens rather than applying a dynamical systems interpretation.  One analysis found that heterothermic endotherms can be broadly categorized into two (mostly) distinct groups: ``daily heterotherms'' and ``hibernators,'' though a few ``intermediate'' organisms fall somewhere between the two groups, including the elephant shrew \cite{geiser,ruf2015daily}. A different statistical analysis found that heterothermic endotherms may fall on a continuum \cite{boyles2011new, boyles2013global}. Here, we show that our selected physiological model of the Arctic ground squirrel, driven by environmental influences, such as photoperiod or temperature, can qualitatively capture body temperature patterns across species, including organisms in Geiser's ``daily heterotherm'' category, ``hibernator'' category, and an ``intermediate'' organism \cite{geiser}. This analysis uses a wide range of tools from applied mathematics to suggest a mechanistic hypothesis in thermal biology: a core thermophysiological mechanism is conserved across heterothermic endotherms, which can respond to and interact with environmental cues.

\section{Results}

\subsection{Model selection, robustness testing, and validation}

To find a thermophysiological model for the Arctic ground squirrel, we employ a recent sparse model selection algorithm designed for partially observed systems \cite{dahsi}. In our system, only one variable, body temperature, is measured (Figure \ref{fig:1}A). To determine the intrinsic dimensionality, we statistically analyze time-delay embeddings of body temperature data and find that the system is two-dimensional (Figure \ref{fig:1}A and Section \ref{dimension}) \cite{kantz2003nonlinear}. Thus, we seek to find a 2-D system of Ordinary Differential Equations (ODE) comprised of monomial terms where the states are the measured temperature, $x$, and a hidden state, $y$, which represents the dynamics of an unknown physiological process. This allows us to find an effective macro-scale physiological model describing the dynamics of coupling between temperature and some unknown internal control process. Sparse model selection evaluates a library of possible terms, $\boldsymbol{{\Theta}}(x,y)$, and composes ODE models by identifying a parsimonious set of terms that match observations of a subset of the system variables (Figure \ref{fig:1}A-B). The search is performed by iteratively estimating the coefficients in the library using variational annealing and applying a threshold as a sparsity penalty. Iterative thresholding on the magnitude of the coefficients for each model term selects terms with large coefficients and sets small coefficients to zero in the next iteration. Of the models that emerge from varying sparsity and increasing model enforcement during optimization, those along the Pareto front balance simplicity and fit to the data, so we select the best fitting eight, nine, and ten active terms for further investigation (Figure \ref{fig:1}B).\footnote{For specific parameters the seven-term model exhibited a limit cycle, but cycling was not robust to small changes to any of the optimized parameters.} These models form a nested family, and the algorithm selects variations on the same model. Our three models of interest can be written in the form 

        \begin{align}
               \dot{x} &= \theta_1 x^2 - \theta_2 x^3 - \theta_3 xy + \theta_4 x^2y - \textcolor{black}{\theta_A y^3} + \textcolor{black}{\theta_B x y^2}, &x(0)= x_0, \label{eqfull1} \\
                 \dot{y} &= \theta_5 x^2 - \theta_6 y^2 - \textcolor{mypink}{\theta_C x^3}+ \textcolor{mypink}{\theta_Dxy}, &y(0)= y_0, \label{eqfull2}
            \end{align}

where the terms in black define the 8-term model, the additional terms appear in the 9-term and 10-term models, and the model coefficients are all assumed to be positive. The nine-term model includes the $\textcolor{mypink}{\theta_C x^3}$ and the ten-term model includes both the $\textcolor{mypink}{\theta_C x^3}$ and the $\textcolor{mypink}{\theta_Dxy}$ terms. However, when we forward simulate these models, the trajectories spike once or not at all, eventually reaching what initially appears to be a steady state. So, further examination of these selected models is necessary. 

The models we found were a good fit to the data using our model selection optimization procedure, but do not immediately exhibit limit cycle behavior at the discovered parameters upon forward simulation. \footnote{Near the parameters estimated during model selection, the trajectories appeared to ``get stuck'' near the origin after spiking once. While the origin initially appeared to be a stable fixed point, further investigation showed that Eqs. (\ref{eqfull1})-(\ref{eqfull2}) exhibits a limit cycle with an infinite period. It is possible to rigorously prove that the system described by Eqs. (\ref{eqfull1})-(\ref{eqfull2}) (with $\theta_C=\theta_D=0$) has an unstable fixed point at the origin by invoking the Chetaev Instability Theorem \cite{chetaev}.} This apparent paradox of robust term selection, which does not produce more globally desired behavior when forward simulated, is unsurprising in light of other known biological models that exhibit limit cycle behavior. Many biological models that exhibit limit cycle behavior often include a scale separation in parameter size (for example, the FitzHugh-Nugamo model \cite{fitzhugh, nagumo} and  Izhikevich's model \cite{izhikevich}, amongst others, have this property) and by thresholding on parameter size, we are removing terms with very small parameters that can be important for generating limit cycle behavior and regulation of the limit cycle period length. Therefore, the discovered models that do not exhibit limit cycles with finite periods are likely missing small terms. 


To enable the models of interest to exhibit limit cycles with a finite period, we introduced a constant term representing a low, basal rate of heat production, $\theta_0$, to the body temperature equation (see Equation (\ref{eqfull1wt})). While metabolic rate and heat production are suppressed during hibernation, some small amount of heat is produced even during torpor to regulate the body temperature, $x$, and ensure the hibernator does not freeze \cite{buck2000effects,wang1979time,laursen2015neuronal}. We found that several low-order terms could be introduced to regulate the period of the limit cycle, but we chose to modify the data-driven models in the simplest way possible. \footnote{Notably, there exists a transformation from our models to Líenard equations when a linear term in $y$ is added to the body temperature equation \cite{lloyd}.} With this simple addition, the models of interest generate stable limit cycles with finite periods. To show that the limit cycle approaches an infinite period with zero leak, we introduced the leak term and simulated the Equations (\ref{eqfull1wt})-(\ref{eqfull2wt}) (with $\theta_C=\theta_D=0$) as $\theta_0 \rightarrow 0$. As expected, this resulted in limit cycle behavior with increasingly long periods, as the phase plane velocity near the origin goes to zero as $\theta_0 \rightarrow 0$. Next, we perform additional robustness testing to determine which structural terms are necessary among the nested family of models. 

\subsection{Term robustness testing}

\begin{figure}[h!]
    \centering
    \includegraphics[width=0.9\textwidth]{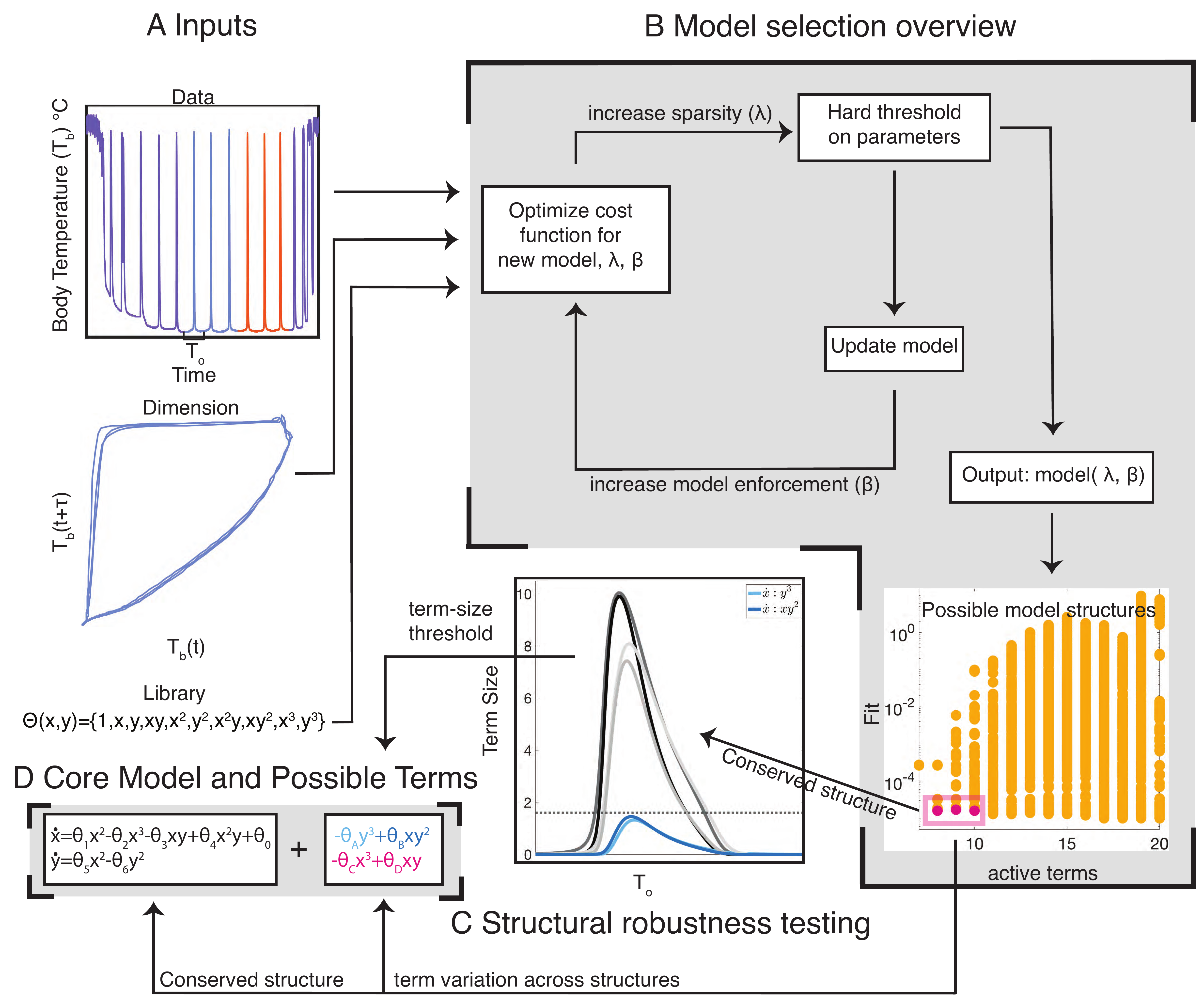}
    \caption{Model selection reveals a seven-term core model and possible higher-order terms that describe experimental observations of Arctic ground squirrel thermophysiology. (A) Inputs for model selection. Data: Body temperature recorded from a free-ranging Arctic ground squirrel \cite{chmura}. Training data is shown in blue. Validation data is shown in orange. Dimension of the system: We compute the dimension of the system using time delay embedding. The system is two-dimensional. We show the phase space reconstruction of the system. Model library: we include monomials up to cubic order in our library. (B) Simplified schematic of the model selection algorithm (for more details, see \cite{dahsi}). Sparse Pareto optimal models with a conserved core structure are shown in pink. (C) A term-size argument is applied to the 8-term model during an interbout arousal to further sparsify our nested family of models by an additional two terms. The terms below the dotted gray line are not needed for oscillation and are therefore placed into the ``possible terms'' category.  (D) Functional form of the core model (black) and additional possible terms (pink and blue). Possible terms in pink emerge as structural variations from the three Pareto optimal models with conserved structure. The terms in blue are eliminated from the core model using the term-size argument.}
    \label{fig:1}
\end{figure}

Here, we test whether the remaining terms are necessary to produce the observed behavior based on overall term size. This is equivalent to keeping only terms that contribute significantly to the dynamics, consistent with dominant balance \cite{white2010asymptotic,tang2023weakident,messenger2021weak}. By forward simulating Equations (\ref{eqfull1wt})-(\ref{eqfull2wt}) with $\theta_C=\theta_D=0$ and examining the term sizes of each component, we find that two additional terms, \textcolor{liteblue}{$\theta_A y^3$} and \textcolor{myblue}{$\theta_B x y^2$}, are negligible across the time-series of a spike (see Figure \ref{fig:1}C).  The terms $\theta_0$, $\theta_5 x^2$, and $\theta_6 y^2$ are excluded from this analysis as they are required to generate finite period limit cycle behavior. In particular, $\theta_5 x^2$ is needed for the existence of a limit cycle, otherwise there is no $x$-dependence in Eq. (\ref{eqfull2wt}). The parameters $\theta_0$ and $\theta_6 y^2$ help to ensure the limit cycle has a finite period by keeping the velocity in the phase plane non-zero near the origin. We now have a larger nested family of models of the form

        \begin{align}
               \dot{x} &= \theta_1 x^2 - \theta_2 x^3 - \theta_3 xy + \theta_4 x^2y + \theta_0 - \textcolor{liteblue}{\theta_A y^3} + \textcolor{myblue}{\theta_B x y^2}, &x(0)= x_0, \label{eqfull1wt} \\
                 \dot{y} &= \theta_5 x^2 - \theta_6 y^2 - \textcolor{mypink}{\theta_C x^3}+ \textcolor{mypink}{\theta_Dxy}, &y(0)= y_0, \label{eqfull2wt}
            \end{align}
where the terms with coefficients $\theta_1,  \theta_2, \theta_3, \theta_4, \theta_5, \theta_6$  (in black), define the core model and the additional terms with coefficients, $\theta_A,  \theta_B, \theta_C, \theta_D$ (blue and pink, Figure \ref{fig:1} D), are non-essential higher-complexity terms. Next, we validate our models using the $l^2$ norm, Akaike information criteria, and dynamic time warping.



\subsection{Model validation}
\label{validation}

\begin{figure}[h!]
    \centering
    \includegraphics[width=0.9\textwidth]{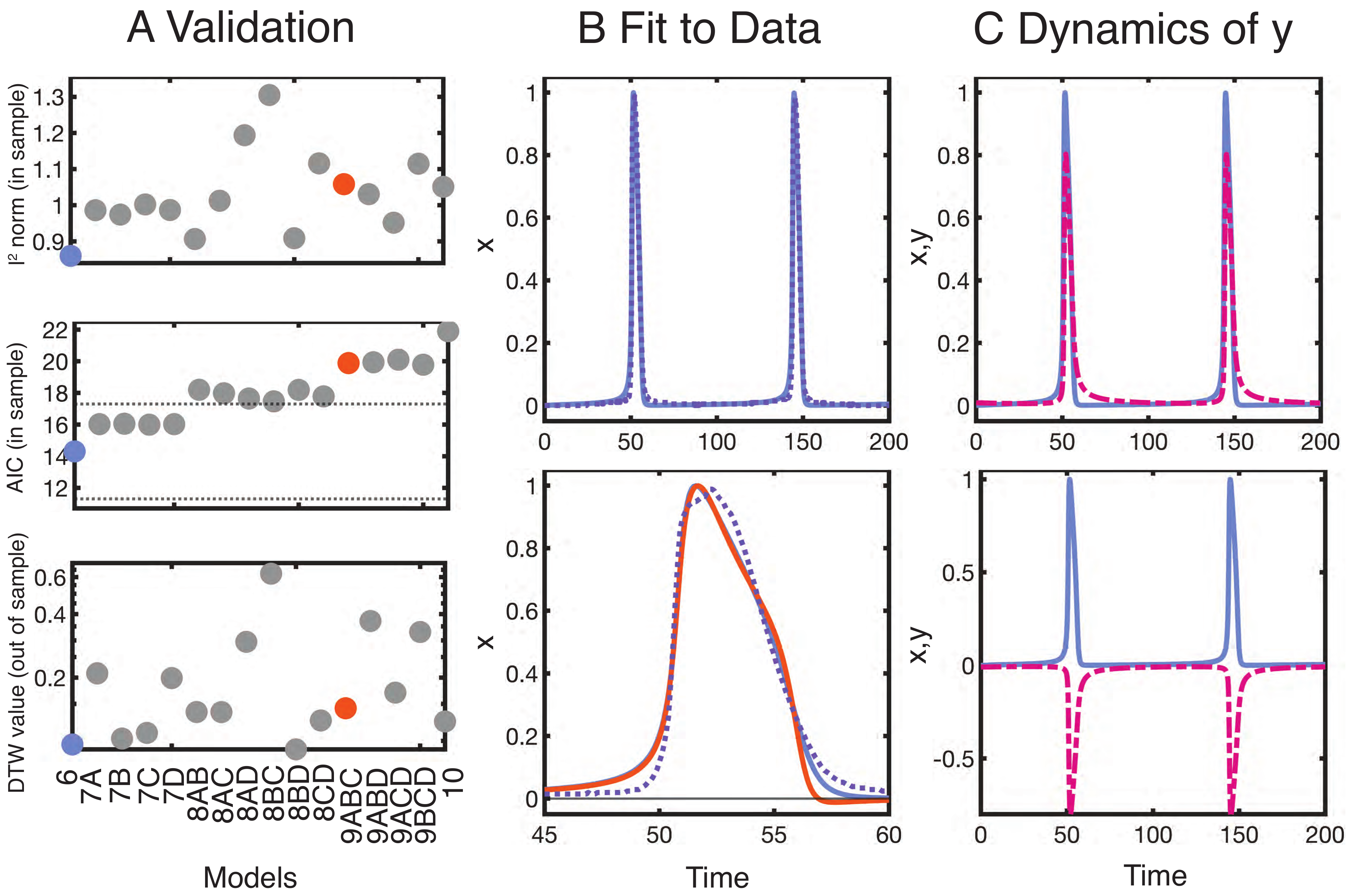}
    \caption{Model validation shows that the simplest model, Model 6, is a reasonable fit to the data. A (top and middle): $l^2$ norm and AIC for the entire nested family of models fit to the first two interbout arousals of the training data. A (bottom): Dynamic time warping values of the model ensemble computed on the validation data. B (top): The fit of Model 6 (blue) to the training data (dotted purple). B (bottom): Close up of the fits during the first interbout arousal for Model 6 (blue) and Model 9ABC (orange), data shown as a dotted purple line. C (top): Dynamics of Model 6 ($x(t)$ in blue, $y(t)$ in pink). The hidden state has a fixed period, lending support to the Hourglass and Threshold Hypothesis \cite{ruf2022}. C (bottom): Introducing a scaling symmetry into the hidden state changes the biological interpretation. In this case, the physiological threshold acts on the difference between multiple metabolites rather than a single metabolite.}
    \label{fig:2}
\end{figure}

To quantitatively evaluate the fit of the model ensemble data to the data, we re-estimate the parameters of each model and compare against the training data (Figure \ref{fig:2}A). We refer to the models by the number of active terms (not including the introduced leak term, $\theta_0$) and the subscript of the parameters of higher complexity terms that the model contains 
(see Eqs. (\ref{eqfull1wt})-(\ref{eqfull2wt})). For example, Model 8AB refers to Eqs. (\ref{eqfull1wt})-(\ref{eqfull2wt}) with $\theta_A \neq 0$, $\theta_B \neq 0$, and  $\theta_C=\theta_D=0$. A complication in parameter estimation and validation for these models is that the interbout arousals of the Arctic ground squirrel are slightly aperiodic, perhaps due to small fluctuations in the decay rate of the unknown metabolite \cite{ruf2022}. To generate a good fit, the forward simulation must match both the height of each body temperature spike and its timing, but a periodic spike will always be misaligned with the aperiodic data. To compensate, the peaks flatten to match most of the non-spiking data, which is near zero. To avoid forcing the data to match multiple periods, we fit all 16 models to the first two spikes of the training data (Figure \ref{fig:2}A). We then evaluate the fit of the models using the $l^2$ norm and Akaike information criteria on the first two spikes of the training data, and validate the models on out-of-sample data, which includes three spikes, using dynamic time warping.

Which model is ``best'' depends on which quantities or qualities we measure. Many of our models are indistinguishable with the current dataset. If we examine the $l^2$ norm between training data and forward simulation without penalizing for model complexity, the models are indistinguishable from each other based on the error (Figure \ref{fig:2}A). If we penalize model complexity using the Akaike information criteria, the simplest Model 6 has the lowest score based on the training data, but most of the seven-term models are statistically indistinguishable (Figure \ref{fig:2}A, relative AIC $\leq 3$ dashed line). Applying dynamic time warping to the out-of-sample data, which roughly compares the shape of the pulse corrected for time-shifting, supports Model 6, Model 7B, Model 7C, and 8BD (Figure \ref{fig:2}A). If we consider more qualitative features of the fit, the simplest model, Model 6, has approximately the correct spike height, but spikes slightly too early and decays too fast (Figure \ref{fig:2}B blue). More complex models, such as 9ABC (Figure \ref{fig:2}B orange), introduce a refractory period after the interbout arousal, which the biological data does not exhibit. Determining which quantitative and subjective (shape of the fit) criteria to use for model selection is difficult without additional experimental information. For simplicity, we continue our analysis with Model 6, which is a reasonable fit to the data but not the only choice. \\

\subsection{Discovered model is consistent with Hourglass and Threshold Hypothesis}

\label{symmetry}

The dynamics of the hidden physiological state, $y$, of Model 6 appear with a constant period rather than a period that varies with body temperature (Figure \ref{fig:2}C). This supports the Hourglass and Threshold Hypothesis over the Torpor Arousal Clock Hypothesis. The ability to differentiate between the two hypotheses is not dependent on the parameterization and structure of the model; the dimensionality of the thermophysiological system alone supports a constant period and the depleted metabolite hypothesis. Planar ODE systems cannot give rise to non-trivial solutions in which $x(t)$ and $y(t)$ have different periods \cite{strogatz2018nonlinear}. However, we can study the identified model's structure, parameterization, and symmetries to extract additional physiological implications. 

Directly relating the hidden state $y(t)$ to a particular metabolite or physiological state is impossible without direct measurements. However, we can computationally explore possible mechanisms and hypothesize about the minimal mechanism of regulation. In the simplest case, we can imagine the hidden state as a molecule that is depleting over time as envisioned by Ruf, Giroud, and Geiser \cite{ruf2022} (see Figure \ref{fig:2}C top). However, we believe it is important to consider all solutions consistent with the data and their interpretations. By introducing a symmetry transformation into the hidden state of the form $\bar{y} = -y$, which does not change the sparsity pattern of the model, we can generate dynamics in the hidden state that become negative (see Figure \ref{fig:2}C bottom), potentially corresponding to the difference between multiple molecules rather than an individual entity. \footnote{Introducing this change of variables to the system results in the the following system \begin{align*}
    \dot{x} &= \theta_1 x^2 - \theta_2 x^3 + \theta_3 x\bar{y} - \theta_4 x^2\bar{y} + \theta_0, &x(0)= x_0,  \\
    \dot{\bar{y}} &= -\theta_5 x^2 + \theta_6 \bar{y}^2, &\bar{y}(0)= \bar{y}_0.
\end{align*}} In this case, it is possible for one of the molecules to increase, as may be expected of a toxic molecule that accumulates during torpor. This alternative parameterization of the model also reflects a fact mentioned in \cite{ruf2022}: there is no way to distinguish between an accumulating metabolite and a depleting metabolite as the driver of interbout arousals. These symmetry transformations do not affect the fit to the measured state, so we cannot distinguish which of these model variants and corresponding metabolic scenarios are correct. Whether representing a single molecule or interactions between multiple molecules, the effective state generates hourglass-like dynamics.

\subsection{Towards a multi-scale model of seasonal physiology}
\label{bif}

Small hibernators exhibit clear seasonal changes in physiology (see Figure \ref{fig:0}). In the summer, the Arctic ground squirrel is homeothermic, with minor fluxuations induced by activity and sleep around a  \textit{physiological fixed point} near 37 $^{\circ}$C, but in hibernation, the organism cannot remain solely at 37 $^{\circ}$C or the low state (near or below 0 degrees Celsius), but must oscillate between the two states to survive. Despite being found using sparse model-selection on only the data during hibernation, our discovered model captures these dynamics, has parameters that have clear physiological meaning, and connects with the known biology of the \textit{circannual clock} \cite{gwinnerbook,shona}. 

Consider the nondimensionalization of Eqs. (\ref{eqfull1wt})-(\ref{eqfull2wt}) with with non-essential terms $\theta_A$=$\theta_B$=$\theta_C$=$\theta_D$=0,


        \begin{align}
               \{\text{dimensionless body temperature}\} ~~\dot{\tilde{x}} &= \omega \tilde{x}^2 - \tilde{x}^3 - \nu (\tilde{x}\tilde{y} -\tilde{x}^2\tilde{y}) + \zeta, &\tilde{x}(0)= \tilde{x}_0, \label{nd3} \\
                 \{\text{dimensionless unknown physiological state}\}~~\dot{\tilde{y}} &= \rho (\tilde{x}^2 -\tilde{y}^2), &\tilde{y}(0)= \tilde{y}_0.  \label{nd4}
            \end{align}

Each of the parameters has a clear physiological interpretation. The parameter $\omega$ controls the amplitude of the oscillations during hibernation. From laboratory experiments, it is known that hibernators attenuate the amplitude of the oscillations in response to environmental factors \cite{maccannell2021elevated}. It was also shown that perturbing temperature during hibernation can alter the timing of reproduction expression but does not interfere with the timing of future hibernation cycles \cite{barnes1990effect}. The parameter $\nu$ controls the seasonally varying coupling between body temperature and the unknown physiological state. $\nu$ is the parameter that controls the transitions from homeothermy to the hibernation thermoregulatory regime and vice versa. Biologically, $\nu$ can be interpreted as a summary output of the circannual clock (see below). The parameters $\zeta$ and $\rho$ influence the period of oscillations during hibernation. $\rho$ controls the scale separation between body temperature and the unknown physiological state and influences the period of oscillations. Biologically, we can interpret $\zeta$ and $\rho$ as modulating the speed of the hourglass and the location of the threshold, perhaps in an environmentally sensitive manner \cite{maccannell2021elevated}.

\begin{figure}[h!]
    \centering
    \includegraphics[width=0.9\textwidth]{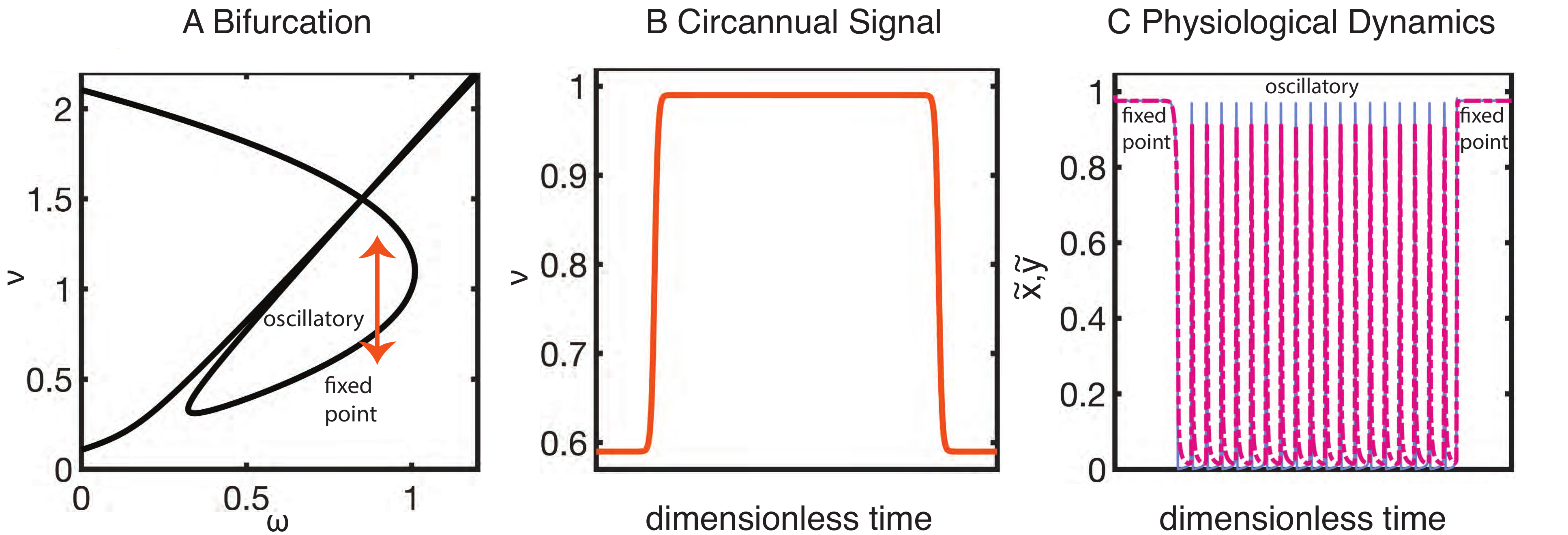}
    \caption{The beginnings of a multi-scale, neuro-physiological understanding of seasonal physiology. A: We find the location of the supercritical Hopf bifurcation in $\omega$-$\nu$-$\zeta$-$\rho$ space and plot Eq. \ref{h} for $\zeta=0.001$ and $\rho=0.1$. This curve divides parameter space into two qualitatively different model behaviors associated with distinct thermoregulatory modes exhibited by the Arctic ground squirrel: a stable fixed point, which corresponds to a regulated body temperature exhibited by the squirrel in the summer, and oscillatory dynamics exhibited during hibernation. Seasonal thermoregulatory changes are associated with the system crossing the boundary that separates fixed point behavior and oscillatory dynamics. B: Biologically, $\nu(t)$ may correspond to a summary output of the circannual clock, a neural mechanism to track seasonal change and generate seasonal decisions. An example trajectory of $\nu(t)$ that allows Eqs. (\ref{nd3})-(\ref{nd4}) to qualitatively capture the body temperature dynamics observed throughout the annual cycle (see Figure \ref{fig:0}, Left). C: The full seasonal physiological dynamics of Eqs. (\ref{nd3})-(\ref{nd4}) as $\nu(t)$ varies as in B. Dimensionless body temperature, $\tilde{x}$, and the dimensionless unknown physiological state, $\tilde{y}$, are shown in blue and purple, respectively.}
    \label{fig:hopf}
\end{figure}

The system undergoes a supercritical Hopf bifurcation as a function of $\nu$, and we can find where this bifurcation occurs in $\omega$-$\nu$-$\zeta$-$\rho$ space. \footnote{It is straightforward to analytically find where this bifurcation occurs in $\omega$-$\nu$-$\zeta$-$\rho$ space. This curve divides parameter space into qualitatively different behaviors and has the form  
\begin{align}
&\nu^4 - 8\nu^3\zeta + 2\nu^3\rho - 4\nu^3\omega - 4\nu^2\rho^2 + 4\nu^2\omega^2 - 8\nu\rho^3 + 16\nu\rho^2\omega- 8\nu\rho\omega^2 - 2\nu^3 + \label{h} \\
    & 36\nu^2\zeta - 6\nu^2\rho + 9\nu^2\omega + 12\nu\rho\omega - 12\nu\omega^2 + 8\rho^3 - 12\rho^2\omega + 4\omega^3 - 54\nu\zeta + 27\zeta =0. \nonumber
\end{align}} The curve is visualized in $\omega$-$\nu$ space for fixed values of $\zeta$ and $\rho$ (see Figure \ref{fig:hopf} A). The transition from a regulated body temperature to an oscillating body temperature during hibernation is associated with a change of stability in the system (see Section \ref{appendix}) and traversing a boundary in parameter space (see Figure \ref{fig:hopf}). Similarly, the exit from hibernation and reversion to a regulated body temperature exhibited in the summer is associated with the opposite change of stability and movement in parameter space (see Section \ref{appendix} and Figure \ref{fig:hopf}A). Assuming $\nu$ is time-dependent, we can find an example trajectory of $\nu(t)$ that allows the system to capture the broad features of seasonal hibernation physiology (Figure \ref{fig:hopf} B and C). 

The dynamics of $\nu(t)$ shown in Figure \ref{fig:hopf} B is qualitatively consistent with our current understanding of the circannual clock. The circannual clock is a neural mechanism that tracks the passage of the year and generates mammalian seasonal decisions, such as hibernation timing, migration timing, mating decisions, and seasonal coat change in organisms like the snowshoe hare and the Arctic fox. The long time scale of these behaviors has made a molecular description of the circannual clock elusive, but some is known about the mechanistic underpinnings of the clock. The clock is thought to be located in the pars tuberalis in the Anterior pituitary gland, and seasonal decisions are thought to be controlled by a population of pars-tuberalis-specific thyrothropic cells, which exhibit a bistable switch in gene expression \cite{shona}. This change causes downstream endocrinological changes that form the basis of seasonal decision-making. At the tissue level, the theorized dynamics of the thyrothropic cells qualitatively resemble the trajectory of $\nu$ over time \cite{shona}. We interpret $\nu$ as a summary output of the circannual clock, and this model as the beginnings of a multi-scale model of seasonal physiology that connects the neurological and physiological scales. 

\subsection{Thermophysiological organization across species}

\label{across}

\begin{figure}[h!]
    \centering
\includegraphics[width=0.9\textwidth]{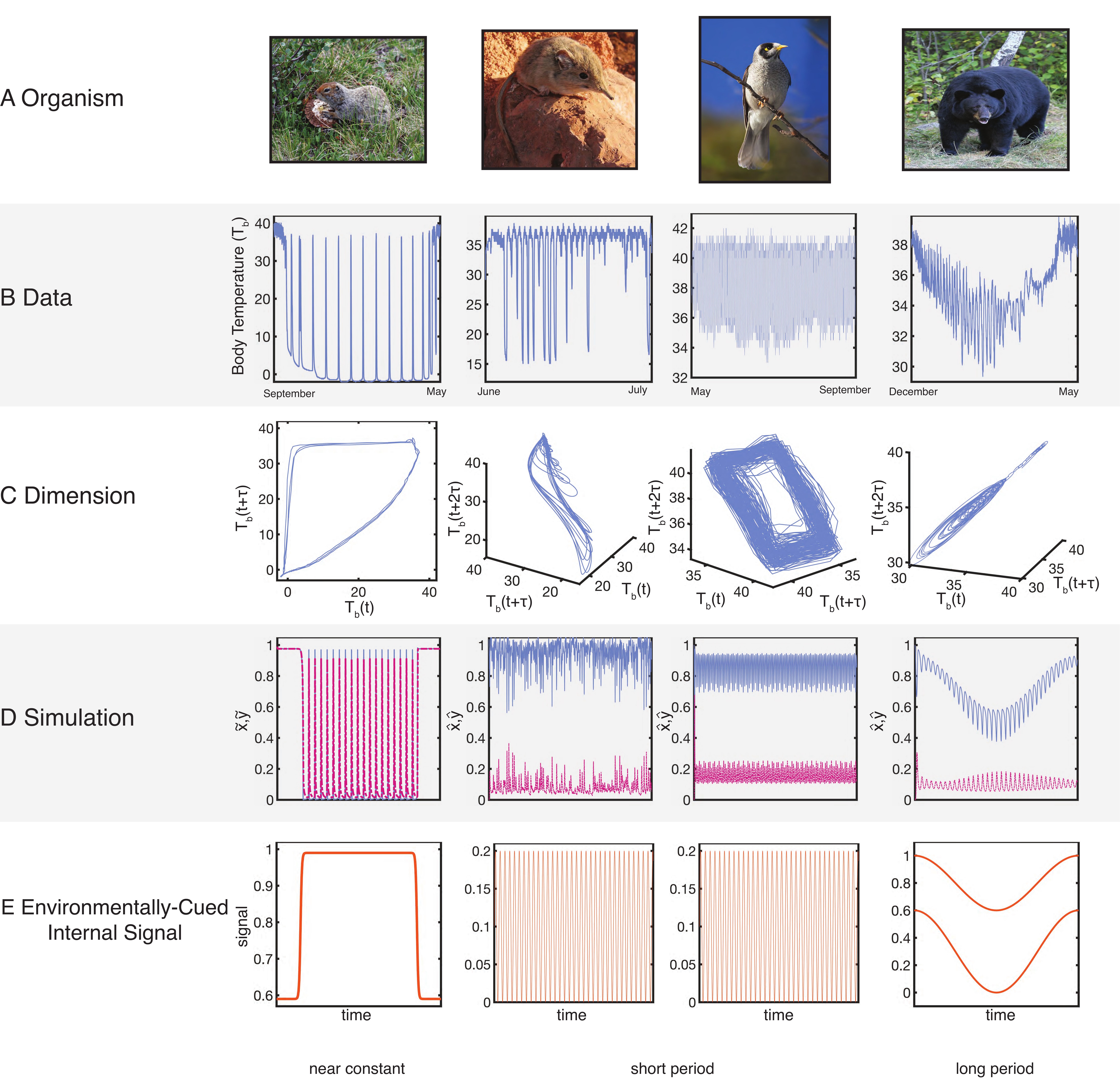}
    \caption{We modify our discovered physiological model of Arctic ground squirrel to include environmental information and find that we can qualitatively match body temperature data recorded from a wide range of species, including a bird, an elephant shrew, and a bear, which also dynamically modulate body temperature. This suggests that a low-dimensional, environmentally sensitive mechanism could regulate body temperature across a diverse range of species. A: Photographs of an Arctic ground squirrel, an elephant shrew, a noisy miner, and a black bear. All images taken from Wikipedia. B: Temperature recordings from an Arctic ground squirrel \cite{chmura}, an elephant shrew \cite{shrew}, a noisy miner \cite{nm}, and a black bear \cite{toien}. C: Phase space reconstructions of underlying thermophysiological systems. These systems are 2-to 3-dimensional. D: Model simulations of body temperature across species (blue) qualitatively match the observed body temperature data in B. Inferred dynamics of a hidden physiological state (pink). E: Internally-encoded environmental information interacting with the core model for each species. Time is dimensionless in panels D and E.}
    \label{fig:3}
\end{figure}

Many species make thermoregulatory switches similar to the Arctic ground squirrel (Figure \ref{fig:3}A), but often the timescale or amplitude of the oscillations is different (Figure \ref{fig:3}B). It is currently unknown if the same thermophysiological system is conserved across species or if different species have fundamentally different systems. Here, we show how our discovered core model can be modified to capture body temperature patterns from a wide range of species, suggesting the possibility of a unified mechanism across species that can respond to environmental signals. 

In the biological literature, there are two analyses that start to understand differences in thermophysiology across species, but both lack mechanistic insights into the possible differences across species. Geiser classified heterothermic endotherms into two groups: ``hibernators'' and ``daily heterotherms'' \cite{geiser}. His rationale for dividing heterothermic endotherms into distinct categories relies on constructing frequency distributions of features of body temperature and other physiological measurements time series data across species \cite{geiser,ruf2015daily}. However, he notes that some species do not clearly fall into either the ``hibernator'' or ``daily heterotherm'' category. In particular, he mentions the elephant shrew as an ``intermediate'' species, which we study here. A different analysis put forward the notion of ``heterothermic-index,'' which is a time series calculation involving body temperature and is a measure of the variability \cite{boyles2011new}. Using this framework, heterothermic endotherms appear to fall along a continuum \cite{boyles2013global}. Both cross-species analyses are valid and examine different features of the time series data, some of which form bimodal distributions, while other features may fall along a continuum, but neither analysis gives a clear physiological or mechanistic explanation of what is different or the same across species. 

Here, we show the possibility of a conserved thermophysiological mechanism by modifying our discovered core model (Equations (\ref{nd_simp1})-(\ref{nd_simp2})) to respond to environmental information. To determine how to add environmental information to our models, we first examine the dimensionality of the thermophysiological system across species using time delay embedding. We find that the dimensionality of the underlying thermophysiological systems for the elephant shrew, noisy miner, and black bear is \textit{relatively} low-dimensional across species (all systems appear to be three-dimensional, see Figure \ref{fig:3}C and Sections \ref{dimension}-\ref{spirit}). To approximate the extra dimensionality, we add environmental information to our core model. Mathematically, these take the form of trigonometric forcing functions. These functions are phenomenological placeholders, not inferred physiological processes. We find that our modified models can \textit{qualitatively} describe the body temperature patterns of ``daily-heterotherm'' species (the noisy miner), an ''intermediate'' species (the elephant shrew), and a large ''hibernator'' (the black bear) (Figure \ref{fig:3}B and D).

To describe the thermoregulatory behavior of the noisy miner and the elephant shrew, we add a time-dependent forcing function with a short period (Figure \ref{fig:3}E). Biologically, this almost certainly corresponds to a circadian signal. As compared to the Arctic ground squirrel, which is shielded from photoperiod (and to some extent thermal information) by its underground burrow, the noisy miner and elephant shrew have direct access to photoperiodic information. Mathematically, this forcing function alters the flow in the phase plane of the system and modifies the limit cycle shape in a temporal manner. This allows the system to be driven by photoperiodic information while still exhibiting spiking behavior.

To capture the thermoregulatory behavior of the black bear, we also add environmental information through forcing functions, which, again, modify the phase plane dynamics and limit cycle shape in a temporal manner. In this case, we find that we must add two time-dependent forcing functions to our model (Figure \ref{fig:3}E, rightmost).  Notably, the period of these forcing functions is much longer than the forcing functions added to the noisy miner and elephant shrew models, which are assumed to be circadian. Biologically, it is plausible that these forcing functions correspond to an entity oscillating on year scale, such as an endocrine molecule with slow, seasonal fluctuations. It is also possible that such an entity could interact with a circadian process, as circadian gene expression was observed in grizzly bears and may also occur in black bears \cite{vincent2023circadian}.

We find we can qualitatively describe a vast range of body temperature patterns across species using a core thermophysiological model modified to respond to different types of environmental information. Small hibernators use just the core mechanism during hibernation with no environmental input, birds and small non-hibernating mammals use the core mechanism modified to include a circadian-dependent signal, and large hibernators use the core mechanism modified to include signals that oscillate with a period much longer than that of the circadian system. Our analysis leads to a more mechanistic understanding of thermophysiology across species, and a new hypothesis: a core thermophysiological mechanism is shared across heterothermic endotherms, which can respond to environmental signals (Figures \ref{fig:3}-\ref{fig:6}). 

\section{Discussion}

\begin{figure}[H]
    \begin{center}
        \includegraphics[scale=0.25]{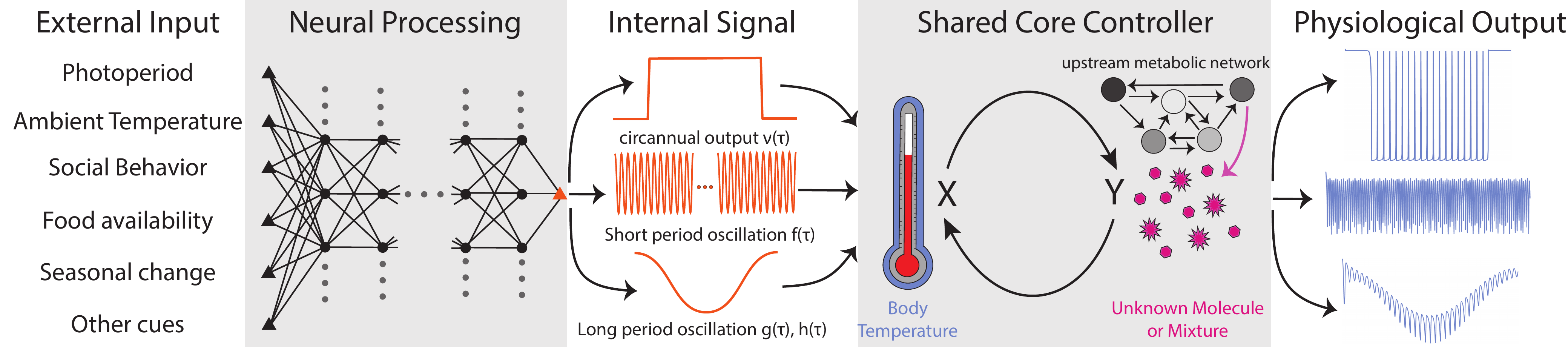}
    \end{center}
    \caption{We hypothesize that a low-dimensional, environmentally sensitive core regulator could control body temperature across a diverse range of species. A broad range of external inputs including photoperiod, ambient temperature, social behavior, food availability, seasonal change, and other cues are neurally-integrated and represented as internal signals, such as the circannual output, a short period or circadian oscillation, or a long period oscillation. The internal signals interact with the core temperature control architecture and produce varied physiological output observed across different species.}
    \label{fig:6}
\end{figure}

Here, we have used a blend of automated model discovery and classical dynamical systems theory to differentiate between two physiological mechanisms hypothesized to drive interbout arousals during small-mammalian hibernation and comment on the possible organization of thermophysiological systems across species using heterothermic endothermy. We find evidence that interbout arousals are driven by hourglass-like dynamics, as hypothesized in \cite{ruf2022} and work towards a multi-scale understanding of seasonal physiological decision making through a bifurcation analysis (see Section \ref{bif}). We modified our discovered core model to include environmental signals and found that we could qualitatively match body temperature recordings from a variety of species, including a bird, an elephant shrew, and a black bear (Figure \ref{fig:3}). This leads to a new hypothesis: a core thermophysiological mechanism is shared across heterothermic endotherms, which can respond to environmental signals (see Figure \ref{fig:6}). 

While the model we find is convincing in demonstrating that a core model could explain body temperature regulation across species, it is not unique. In addition to other structures that arise from symmetry transformations (see Section \ref{symmetry}), there are likely other dynamical systems models that could adequately describe the thermophysiology of the Arctic ground squirrel. Some models that emerge from our model selection procedure do not immediately exhibit limit cycle behavior upon forward simulation. In these cases, it was difficult to know which terms need to be added ``back'' to the model to generate a limit cycle, or, in the case of poor fits (see Figure \ref{fig:1}B, far off the Pareto front), if a limit cycle was even possible. Recall that there is no definitive technique for ruling out limit cycle behavior for general dynamical systems models, though it is possible in special cases (for examples of such systems see \cite{strogatz2018nonlinear}). Mathematically, it would be of interest to determine if it is possible to rule out limit cycle behavior for any broader class of planar Ordinary Differential Equations and study additional model structures consistent with this data. To limit the computational burden of the initial model selection and given the fact that we cannot determine determine the full class of models that can describe the thermophysiology of the Arctic ground squirrel for a single data set, we did not attempt to understand the possible variety of models that emerge if we examined other body temperature data sets from different Arctic ground squirrels. We successfully found and analyzed a data-driven model of Arctic ground squirrel thermophysiology, but future work will likely find alternative or modified model structures that can produce similar behavior. However, we do expect our discovered model to generalize well across data from distinct individuals. Body temperature dynamics across individual squirrels are remarkably similar (see Figure 1 in \cite{wilbur2022survival}), and the existing subtle variations could be easily captured by different parameterizations of Eqs. (\ref{nd3})-(\ref{nd4}). 

Modifying our sparse model selection algorithm for partially observed systems is an exciting future direction of research at the intersection of optimization and dynamical systems theory. We hope to improve two features of the algorithm raised by this study: threshold design and temporal dynamics enforcement. Here, we used a threshold on coefficient size to promote sparsity in our models, but this caused issues for models with a scale separation in the parameter coefficients, such as Model 6. Note that in Model 6, $\theta_0$ is much smaller than many of the parameters in the model and therefore will always be ``cut'' out. Other approaches use a threshold on term-size \cite{tang2023weakident,messenger2021weak}, but it is not obvious that thresholding on term-size would have been superior in our case, as $\theta_0$ may have also had a minute term-size relative to all other terms in the model across all time, given that $\theta_0<<1$. Designing a threshold to promote sparsity and respect scale separation in both coefficient size and term size is a challenging problem in model selection for partially observed systems. With respect to temporal dynamics enforcement, the approach we used enforces the dynamics locally rather than globally in time through the discretization \cite{dahsi}. Effectively, the model must fit the data, five samples at a time, rolling across the entire data set. This means longer-term consistency with multiple oscillations is not enforced via forward simulation during model selection. Enforcing the dynamics globally in time throughout model selection is computationally costly and often not practical, as broad swaths of the parameter space will lead to models that are either unstable or diverge when forward simulated, given our library of nonlinear monomials. Striking the correct balance of local and global enforcement of dynamics in time during model selection is another challenging problem where connections between dynamical systems theory and optimization may be fruitful. 


Beyond the algorithmic design, we made several practical modeling choices that should be noted. We included monomials up to order three in our library during model selection. We drew inspiration from classical dynamical systems models from neuroscience, which typically include terms up to third degree, such as the FitzHugh-Nagumo Model \cite{fitzhugh,nagumo}. Models with higher-order terms, such as quartic and quintic terms, may work as well or better but would be more computationally challenging to select. Another choice was which term to add to the selected models to constrain the period of the limit cycle. We chose to add a small constant to the body temperature equation as the simplest option, but other small linear terms would also regulate the period length of the limit cycle. In future work, we wish to understand if our models can be transformed into Líenard systems or if we have discovered any candidates from a broader and distinct class of oscillatory models. While we evaluated the fit of our models through multiple lenses, there is always the risk of overfitting the parameters, given the limited oscillatory data we used for training and validation. Given this challenge, it is surprising that we were able to discover a model that captures the oscillatory Arctic ground squirrel body temperature data (Section \ref{validation}), capture broad seasonal transitions between thermoregulatory modes (Section \ref{bif}), and generalizes across multiple different species (Section \ref{across}). In the qualitative modeling, we did not examine all possible ways to add environmental information. We demonstrate the existence of one version of a core model and a suite of extended, environmentally driven models, but we do not expect uniqueness. Similarly, it is important to note that the hidden state in our models of thermophysiology across species may not correspond to the same underlying biological entity in each species. We simply claim that the thermophysiological systems may have core components that interact in the same fashion across species, whether due to genetic conservation or convergence, both of which are plausible.

Despite these limitations and caveats, our discovered macro-scale physiological model can be used to guide the ongoing search for molecular mechanisms that underpin small mammalian hibernation physiology based on the dynamics of the hidden state. Due to a scaling symmetry in our model, we cannot differentiate between an accumulating or depleting entity, or whether the driver is a single entity or an amalgamation of multiple entities (see Section \ref{symmetry}). In practice, the causative agent may take the form of a single metabolite, the ratio of key metabolites, the difference between key metabolites, or something else entirely. Our model of hibernation physiology would be validated if the currently unknown causative agent oscillates with the same period as body temperature dynamics during hibernation and is constant during summer, when the organism exhibits a stable body temperature, or invalidated if not. A key point is that the causative agent may not be low in the summer. It may be relatively high, but we expect it to be constant. Functionally, experimental tests of our model would take the form of metabolomic or endocrine sampling over the course of the annual cycle and a demonstration of the causative nature of the discovered entity on interbout arousals. It is important to note that our qualitative modeling presented in see Section \ref{bif} captures only the broad features of the seasonal body temperature time series and was not intended to model the ``flickering'' dynamics that occur at the entrance and exit of hibernation \cite{oro2021}. In future work, we will modify the model to better capture these dynamics, but this was not the intention of this modeling effort.


We discover a new data-driven physiological model in the style of the Fitzhugh-Nagumo model \cite{fitzhugh,nagumo}, the NPZ model \cite{franks1986behavior}, and the Brusselator model \cite{lefever1971chemical,prigogine1967dissipative,prigogine1968symmetry}, which distill the incredible complexity and retain only the essential features of the underlying system. We used our discovered model to provide new insights into hibernation physiology, posit the beginnings of a multi-scale model of hibernation that spans the neurological and physiological scales, and put forward a hypothesis in thermal biology: a core thermophysiological mechanism is shared across heterothermic endotherms, which can respond to environmental signals (see Figure \ref{fig:6}). Our analysis starts to describe how physiological systems are affected and controlled by the external environment, and as more physiology is uncovered, we can further refine our models. In the future, we hope to further understand which features of the external environment, such as photoperiod, ambient temperature (see Figure 2 in \cite{nowack}), magnetic fields, social interactions, and food availability, influence thermophysiology and physiology more broadly in potentially nonlinear and seasonally varying ways. 



\section{Methods}

\subsection{Data sets}

We used data sets from the following papers: \cite{chmura,toien,miner,shrew}. If the data were not available with the paper, we reached out to the authors directly.

\subsection{Dimension calculation used time delay embedding}

\label{dimension}

Our goal in using time delay embedding was to determine an approximate dimension for each of the thermophysiological systems. To do so, we applied MATLAB's \texttt{phaseSpaceReconstruction()} function to time series data from \cite{chmura,toien,miner,shrew} using default settings. No smoothing was applied to the Arctic ground squirrel data. We used MATLAB's \texttt{smoothdata()} function to smooth the noisy miner time series (``loess'' option) and the elephant shrew time series (``gaussian'' option) (see Figure \ref{fig:datasmooth}). It is important to note that we digitized the black bear time series using \texttt{\url{https://automeris.io/}} \cite{WebPlotDigitizer}. This process resulted in a data set with non-unique values. We added a small random value of size 1x10$^{-10}$ to each body temperature data point to slightly scatter the data to unique values.  Then, we applied MATLAB's \texttt{interp1()} function to generate higher-frequency data, which requires unique points. We then applied MATLAB's \texttt{smoothdata()} function (``loess'' option) to the interpolated data (see Figure \ref{fig:datasmooth}). To compute the dimension of the Arctic ground squirrel thermophysiology system, we used the segment of data used during model selection (not including the validation portion). For all other dimension calculations, we computed the dimension using a smoothed version of the full time series.

\subsection{A note on the spirit of these calculations}

\label{spirit}

It should be noted that the dimension calculation is sensitive to changes in the data and the data-smoothing technique applied. It is also likely that the interpolation we performed on the black bear data had some impact on the dimension. However, our goal is not to definitively state the dimension of the effective thermophysiology system for any given species, but rather to get an understanding of coarse dimensionality (are the thermophysiological systems relatively low-dimensional (2-5 dimensional) or relatively high-dimensional (10-100 dimensional)?). Across species, we find that the dimensionality of thermophysiological systems is relatively low and can be qualitatively described using the models we lay out here, but that is not to say that alternative choices would not have slightly changed the dimensionality. Additionally, changing hyperparameters of MATLAB's \texttt{phaseSpaceReconstruction()} would likely also change the dimensionality slightly. The ultimate test of dimensionality cannot be found in the statistics of an embedding algorithm or the associated hyperparameters of the algorithm. The ultimate test of dimensionality is whether the model can capture the observation. We were able to show this in a quantitative manner for the Arctic ground squirrel and qualitatively for all other species we examined. 

\subsection{Model selection}

\begin{figure}[H]
    \centering
    \includegraphics[width=0.9\textwidth]{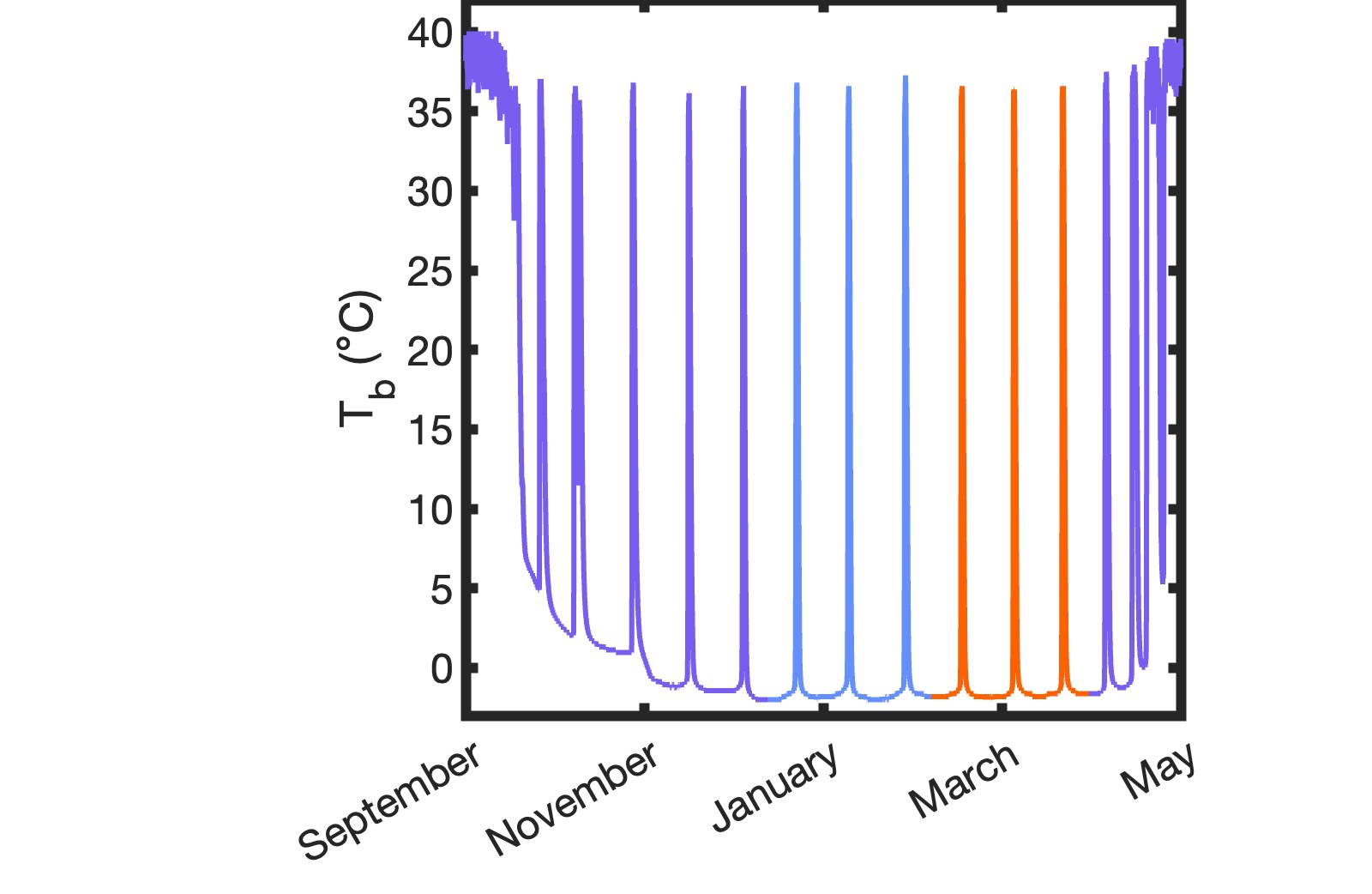}
    \caption{Zoom in of data in Figure \ref{fig:1}. Data: Body temperature recorded from a free-ranging Arctic ground squirrel \cite{chmura}. Training data is shown in blue. Validation data is shown in orange.}
    \label{fig:data}
\end{figure}

To perform model selection on Arctic ground squirrel body temperature data, we used the data found in 9903\_M\_Toolik\_temp.csv in the supplement of \cite{chmura}. We opted to use a segment of the Arctic ground squirrel body temperature data set that was six spikes long. Specifically, we used the data from index 10811 to index 13747 as training data and from index 13748 to index 16573 as validation data (see Figures \ref{fig:1}A and \ref{fig:data}). Including more data was infeasible as doing so would increase the run time of the algorithm, which in some cases took up to 48 hours on an HPC cluster. We applied min-max scaling to the training and validation data. We note that using different data sets may produce different model selection results (see Discussion for further information on modeling choices and limitations). 

Our model selection library includes monomials up to the cubic degree. We ran the algorithm with 100 seeds on the Quest High-Performance Computing Cluster at Northwestern University. One seed did not finish after 48 hours, so we consider the results from the 99 seeds that did finish. The algorithm has two main hyperparameters: $\lambda$ and $\beta$, which define the hard-threshold value and the exponent of the model error weight, respectively. Note that the model error weight is defined as $R_f = R_{f_0}\alpha^\beta.$ $\lambda$ ranged from 0.05 to 2.0 and was incremented by 0.05. $\beta$ ranged from 0 to 120 and was incremented by 1. We examined action vs. $\beta$ plots and decided to look at selected models for $75 \leq \beta \leq 85$ (see Figure \ref{fig:betaaction}). We were looking for a range in $\beta$ such that the action vs. $\beta$ curve plateaued, which indicates an appropriate balance of data error and model error enforcement. For $\beta>85$, the model error seemed to be over-enforced, leading to curves that increased after the plateau. We did not consider selected models for $\beta<75,$ as many action vs. $\beta$ curves were still increasing, indicating not enough model error enforcement. 

\subsection{Identifiability}

To ensure that the entire suite of models defined by Eqs. (\ref{eqfull1})-(\ref{eqfull2}) were structurally identifiable ahead of parameter estimation, we used \texttt{StructuralIdentifiability.jl} \cite{structidjl} to check that Model 10 was identifiable after setting $\theta_4$ to its numerical value found during model selection (specifically the value of $\theta_4$ for the eight-term model recovered during model selection, though the values of $\theta_4$ in the nine-term and ten-term models are almost identical). We found that all parameters were globally identifiable with $\theta_4$ fixed. If Model 10 is identifiable, then our other models of interest, which are subsets of Model 10, will be as well. It should be noted that there is no identifiability guarantee for the models we found during model selection \cite{massonis2023distilling}. 

\subsection{Parameter estimation on selected models}

We performed parameter estimation for the kinetic parameters and the initial condition of the hidden state across our suite of models of interest using LBFGSB \cite{zhu1997algorithm}. The upper and lower bounds were chosen to be 50\% and 150\% of the optimized parameters discovered during model selection, respectively. For the initial condition of the hidden state and leak term, we allowed prescribed upper and lower bounds that let each be close to zero. For the initial condition of the hidden state, we allowed the bounds to explore both negative and positive values. Across the 16 models, we used 10,000 seeds sampled using Latin Hypercube Sampling and scaled between the lower and upper bounds as implemented here \cite{urquhart_surrogate-based_2020}. We used an $l^2$ norm on the measured state as the cost function. When we refit across the ensemble, we used the parameter set with the lowest cost function value regardless of the convergence flag status. To analyze the sensitivity of the fit, we examined the Hessian of the cost function at the best fit for Model 6. As expected, the stiffest directions are controlled by $\theta_0$ and $\hat{y}(0)$ (see Supplemental Figures \ref{fig:sloppy1}). Subtle changes in these parameters will cause the limit cycle to be lost and, thus, will result in a poor fit. Forward simulations of the model were done using \texttt{OrdinaryDiffEq.jl}, a component package of \texttt{DifferentialEquations.jl} \cite{rackauckas2017differentialequations}. We used the \texttt{Optimization.jl} implementation of LBFGSB \cite{vaibhav_kumar_dixit_2023_7738525}. Other helper packages were also used, including: \texttt{Random.jl}, \texttt{LinearAlgebra.jl}, \texttt{DataFrames.jl}, \texttt{CSV.jl},  \texttt{SciMLSensitivity.jl} \cite{rackauckas2020universal}, \texttt{ForwardDiff.jl} \cite{RevelsLubinPapamarkou2016}, and \texttt{ModelingToolkit.jl} \cite{ma2021modelingtoolkit}.

\subsection{Code}

Code used for this analysis can be found here \url{https://github.com/cefitzg/heterothermy.git}.

\section{Appendix}

\subsection{Local stability analysis}
\label{appendix}

The Arctic ground squirrel thermophysiology model we discovered 
(Eqs. \ref{eqfull1wt}-\ref{eqfull2wt}) 
can also describe seasonal transitions between summer homeothermy and the hibernation thermoregulatory regime. In the summer, the Arctic ground squirrel is homeothermic, with minor fluctuations induced by activity and sleep around a  \textit{physiological fixed point} near 37 $^{\circ}$C. During hibernation, the organism cannot remain solely at 37 $^{\circ}$C or the low state (near or below 0 $\circ$ C), but must oscillate between the two states to survive. Therefore, any physiological fixed points corresponding to high-temperature or low-temperature states must be unstable (or non-attractive). Mathematically, we expect the stability of the model's high-temperature fixed point to transition from stable to unstable in the fall and back to stable in the spring. The model's low-temperature fixed point should always be unstable, as staying at a low temperature without oscillations at any time of year would be fatal for the organism. To confirm this, we perform a stability analysis of our model. 

To perform a stability analysis, we perform a second nondimensionalization on Eqs. (\ref{eqfull1wt})-(\ref{eqfull2wt})) with non-essential terms $\theta_A$=$\theta_B$=$\theta_C$=$\theta_D$=0. The simplified nondimensionalized model has the form

               \begin{align}
                 \{\text{dimensionless body temperature}\} ~~ \dot{\hat{x}} &= \hat{x}^2 -\hat{x}^3-\gamma\hat{x}\hat{y} + \xi\hat{x}^2\hat{y}  + \iota, \hat{x}(0)= \hat{x}_0, \label{nda} \\
                 \{\text{dimensionless unknown physiological state}\}~~\dot{\hat{y}} &= \kappa \left( \hat{x}^2 - \hat{y}^2 \right), \hat{y}(0)= \hat{y}_0. \label{ndb} 
            \end{align}

If we assume $\gamma \approx \xi$, which is true of the parameters estimated during validation (see Section \ref{nd}), Eqs. (\ref{nda})-(\ref{ndb}) has the following simplified form

               \begin{align}
                \{\text{dimensionless body temperature}\} ~~ \dot{\hat{x}} &= \hat{x}\left(\hat{x}-\gamma \hat{y}\right) \left( 
                1 - \hat{x} \right) + \iota, \hat{x}(0)= \hat{x}_0, \label{nd_simp1}\\
                \{\text{dimensionless unknown physiological state}\}~~\dot{\hat{y}} &= \kappa \left( \hat{x}^2 - \hat{y}^2 \right), \hat{y}(0)= \hat{y}_0. \label{nd_simp2} 
            \end{align}

The model parameters of Eqs. (\ref{nd_simp1})-(\ref{nd_simp2}) has interpretable physiological meaning. $\gamma$ describes the seasonal coupling between body temperature and the unknown physiological state. The parameters $\iota$ and $\kappa$ influence the period of oscillations during hibernation. Eqs. (\ref{nd_simp1})-(\ref{nd_simp2}) has three fixed points: $(\hat{x},\hat{y}) = (0,0)$, $(1,-1)$, and $(1,1)$ when $\iota = 0.$\footnote{It is reasonable to take $\iota=0$ to examine fixed points sufficiently far away from the origin, as the higher order terms will locally dominate $\iota$, as $\iota<<1$, but this is not reasonable near the origin where $\iota$ dominates. So under the $\iota=0$ approximation, we cannot comment on the stability of the fixed point at the origin, as our assumption does not hold. Additionally, note that the fixed point $(\hat{x},\hat{y}) = (1,-1)$ is unstable for all $\gamma$ and $\kappa$.} A linear stability analysis of Eqs. (\ref{nd_simp1})-(\ref{nd_simp2}) when $\iota=0$ shows that $(\hat{x}, \hat{y})=(1,1)$, which corresponds to the high-temperature state, is stable for $0<\gamma <1$ and is an unstable saddle for $\gamma>1$. For $\gamma=1$, linear stability analysis is indeterminate. Biologically, the thermoregulatory transitions between homeothermy and the hibernation thermoregulatory regime can be understood as seasonal variation in $\gamma$, perhaps driven by the circannual clock, which corresponds mathematically to a seasonal change of stability of the fixed point $(\hat{x},\hat{y}) = (1,1)$, the high-temperature fixed point. While the Arctic ground squirrel is homeothermic during the summer ($\gamma < 1$), the thermophysiological system stays near $(\hat{x},\hat{y}) = (1,1)$ with minor fluctuations. During hibernation ($\gamma > 1$), $(\hat{x},\hat{y}) = (1,1)$ is unstable, and the system snaps to the limit cycle, which describes the the hibernation thermoregulatory regime regime. Thus, the stability of the point near (1,1) is controlled by $\gamma$ in a seasonal manner as expected. This seasonal transition and change of stability are shown in Figure \ref{fig:hopf} for the same model with a different choice of nondimensionalization. 

Next, we check the stability of fixed points near the origin, which must always be unstable from a biological perspective, as an organism remaining in the low state at any time of year for an indefinite period would die. We neglect the full model's cubic terms, as they are negligible for $\hat{x} \approx \hat{y} \approx 0$, and examine the system              \begin{align*}
                \dot{\hat{x}} &= \hat{x}\left(\hat{x}-\gamma \hat{y}\right) + \iota, \hat{x}(0)= \hat{x}_0, \\
                \dot{\hat{y}} &= \kappa \left( \hat{x}^2 - \hat{y}^2 \right), \hat{y}(0)= \hat{y}_0.\\
            \end{align*} We restrict ourselves to examining only the real fixed point in the positive quadrant, which has the form $(\hat{x}, \hat{y}) = \left(\sqrt{\dfrac{-\iota}{(1-\gamma)}}, \sqrt{\dfrac{-\iota}{(1-\gamma)}} \right).$ For $\gamma$, $\iota > 1$, there exists a real fixed point near the origin. One can show that the origin will be unstable if 
             \begin{align}
                &1<\gamma<2(1+\kappa) \label{ineq1}\\
                &(\gamma + 2\kappa - 2)^2\sqrt{\dfrac{-\iota}{1-\gamma}} - 8\gamma\kappa < 0. \label{ineq2}
            \end{align}
           Notice that the second inequality will be dominated by the second term, $- 8\gamma\kappa$,  as $\iota << 1$ in our simulations. This leads to complex eigenvalues with positive real parts and, therefore, an unstable equilibrium near the origin. As expected, the stability of $(\hat{x},\hat{y}) = (1,1)$ varies in a seasonal manner and $(\hat{x}, \hat{y}) = \left(\sqrt{\dfrac{-\iota}{(1-\gamma)}}, \sqrt{\dfrac{-\iota}{(1-\gamma)}} \right)$ is always unstable in the parameter regime described by the Inequalities (\ref{ineq1})-(\ref{ineq2}).

\subsection{Supporting Figures}



\begin{figure}[H]
    \centering
    \includegraphics[width=0.9\textwidth]{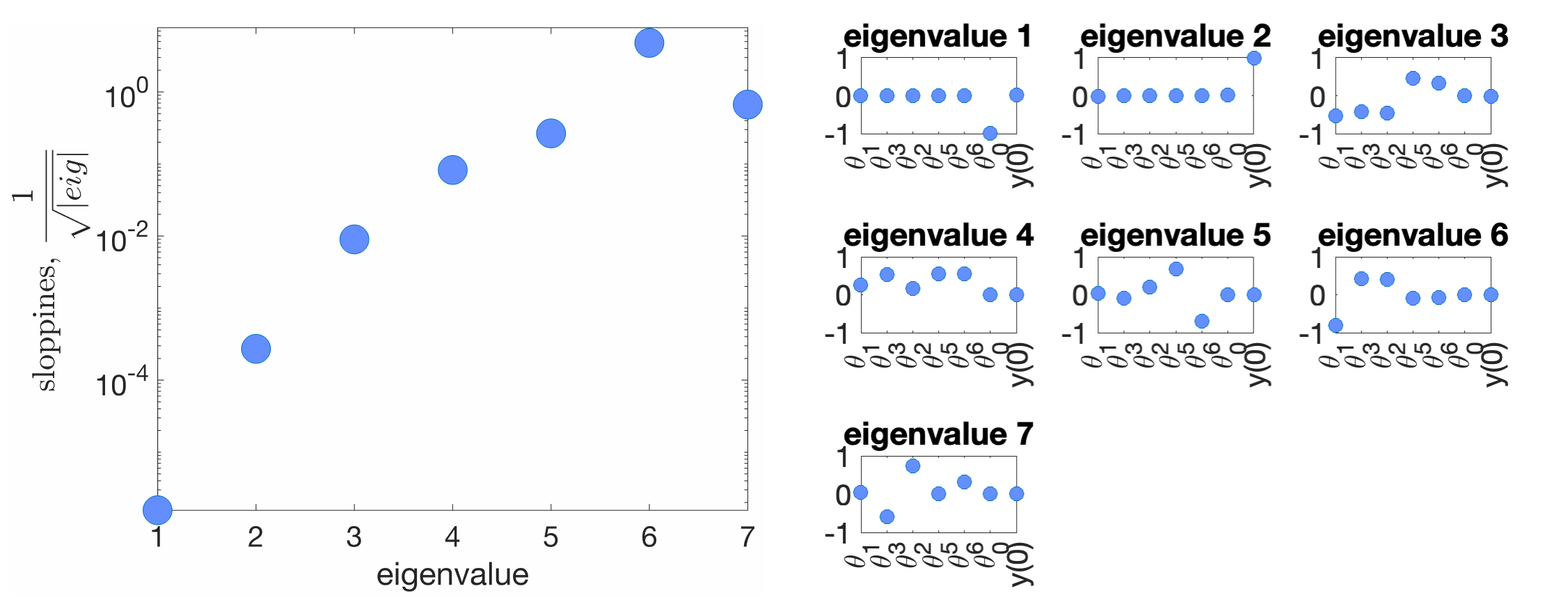}
    \caption{Left: The eigenvalue spectra of the Hessian of the cost function for Model 6 at the best fit exhibit sloppiness (see \cite{gutenkunst2007universally}). Right: Eigenvector contributions from each model parameter (see \cite{gutenkunst2007universally}). As expected, $\theta_0$ and $y(0)$ define the stiffest directions.}
    \label{fig:sloppy1}
\end{figure}

\begin{figure}[H]
    \centering
    \includegraphics[width=1.0\textwidth]{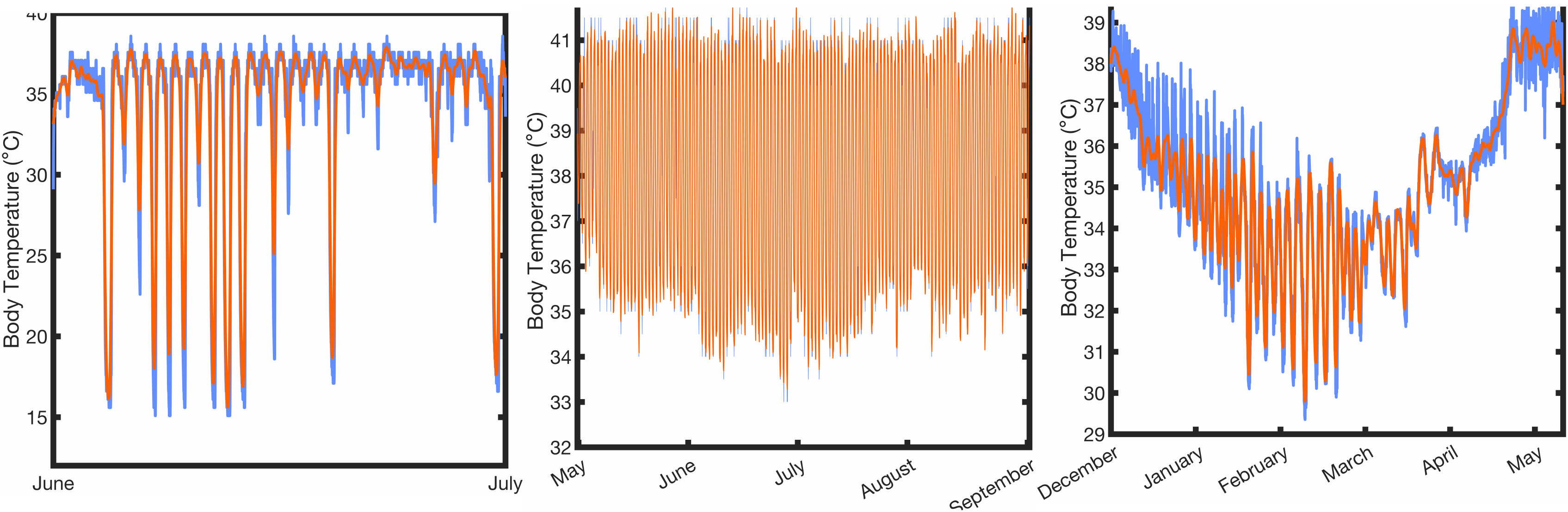}
    \caption{Original data in blue and smoothing applied to the data in orange for the elephant shrew \cite{shrew}, the noisy miner \cite{nm}, and the black bear \cite{toien}.}
    \label{fig:datasmooth}
\end{figure}

\begin{figure}[H]
    \centering
    \includegraphics[scale=0.15]{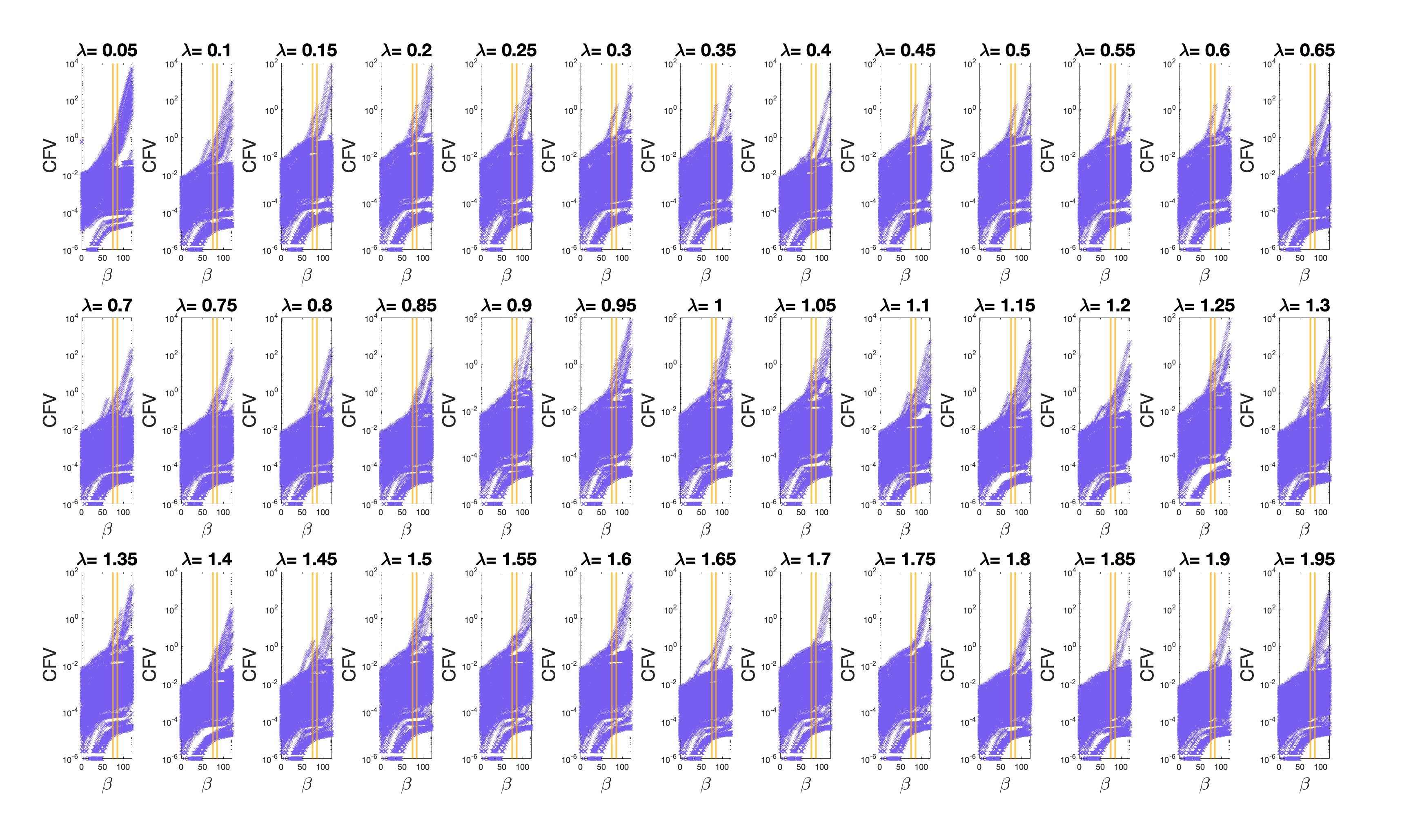}
    \caption{Cost function value (also known as ``action'') vs. $\beta$ plots at different sparsity thresholds (purple). An action vs. $\beta$ curve that is plateaued indicates an appropriate balance of data error and model error enforcement. For this reason, we examined models for $75 \leq \beta \leq 85$ in our study (cutoffs shown as yellow lines). }
    \label{fig:betaaction}
\end{figure}



\subsection{Note on Figure \ref{fig:3}}

We use photographs from Wikipedia of the Arctic ground squirrel \cite{ags_pic}, the elephant shrew \cite{es_pic}, the noisy miner \cite{nm_pic}, and the black bear \cite{bb_pic}.

\subsection{Nondimensionalization and assumptions on model parameters}

\label{nd}

Consider Model 6, which has the form

        \begin{align*}
               \dot{x} &= \theta_1 x^2 - \theta_2 x^3 - \theta_3 xy + \theta_4 x^2y + \theta_0, &x(0)= x_0, \\
                 \dot{y} &= \theta_5 x^2 - \theta_6 y^2, &y(0)= y_0.
        \end{align*}

We can nondimensionalize the system by introducing the scaling, $\hat{x} = \dfrac{x}{s_1}$, $\hat{y} = \dfrac{y}{s_2}$, and $\tau = \dfrac{t}{s_0}$ with $s_1=\dfrac{\theta_1}{\theta_2}$, $s_2=\dfrac{\sqrt{\dfrac{\theta_5}{\theta_6}} \theta_1}{\theta_2}$, and $s_0=\dfrac{\theta_2}{(\theta_1)^2}$. We assume $\theta_i>0$ for $i=0,\dots,6.$ The nondimensionalized system has the form

               \begin{align*}
               \dot{\hat{x}} &= \hat{x}^2 -\hat{x}^3-\gamma\hat{x}\hat{y} + \xi\hat{x}^2\hat{y}  + \iota, \hat{x}(0)= \hat{x}_0, \\
                \dot{\hat{y}} &= \kappa \left( \hat{x}^2 - \hat{y}^2 \right), \hat{y}(0)= \hat{y}_0,
            \end{align*}

            where $\gamma=\dfrac{\sqrt{\dfrac{\theta_5}{\theta_6}}\theta_3}{\theta_1}$, $\xi=\dfrac{\sqrt{\dfrac{\theta_5}{\theta_6}}\theta_4}{\theta_2}$, $\iota = \dfrac{\theta_0 (\theta_2)^2}{(\theta_1)^3}$, and $\kappa = \dfrac{\sqrt{\theta_5 \theta_6}}{\theta_1}$.

            We make the assumption $\gamma \approx \xi$. During parameter estimation, we fix $\theta_4$ to make the models identifiable ($\theta_4=9.99995$, the value of Model 8AB that came out of model selection with $\theta_0=0$). After reoptimization, the numerical values of the nondimensionalized parameters are $\gamma=0.9930155091$, $\xi=0.9776338555$, $\iota=0.00002991498959$, and $\kappa = 0.2621331463$. From this, we can see that $\gamma \approx \xi$. Under this assumption, we can factor the system, which yields the model we analyze in Section \ref{appendix}. 

               \begin{align*}
                \dot{\hat{x}} &= \hat{x}\left(\hat{x}-\gamma \hat{y}\right) \left( 
                1 - \hat{x} \right) + \iota, \hat{x}(0)= \hat{x}_0,\\
               \dot{\hat{y}} &= \kappa \left( \hat{x}^2 - \hat{y}^2 \right), \hat{y}(0)= \hat{y}_0.
            \end{align*}

In Section \ref{bif}, we used the following scalings  $\hat{x} = \dfrac{x}{s_1}$, $\hat{y} = \dfrac{y}{s_2}$, and $\tau = \dfrac{t}{s_0}$ with $s_1=\dfrac{\theta_3}{\theta_4}$, $s_2=\dfrac{\sqrt{\theta_5\theta_6} \theta_3}{\theta_6 \theta_4}$, and $s_0=\dfrac{(\theta_4)^2}{(\theta_3)^2\theta_2}$ to find the dimensionless system described by Eqs. (\ref{nd3})-(\ref{nd4}).

\subsection{Modified model functional forms and parameter values}

\subsubsection{The Arctic ground squirrel model}

In Figures \ref{fig:hopf} B and C and \ref{fig:3} D and E, we simulate Eqs. (\ref{nd3})-(\ref{nd4}) using the parameter values $\omega = 0.99,$ $\rho = 0.1,$ $\iota = 1$ x 10$^{-4}$, $\tilde{x}_0=0.99$, and $\tilde{y}_0 =0.99$. We take

\begin{equation*}
\nu(t)=
\begin{cases}
0.2(\text{tanh}(0.005(\tau-3,500)))+0.79      & \text{if } 0\leq \tau \leq 12,000, \\
0.2(1-\text{tanh}(0.005(\tau-20,500)))+0.79 & \text{if } 12,000 < \tau \leq 24,000.
\end{cases}
\end{equation*}

\subsubsection{The noisy miner model}


Consider a variant of the Arctic ground squirrel model with slight changes to the parameters and a single external forcing function, $f(\tau)$:
                \begin{align}
                \dot{\hat{x}} &= -\left(\hat{x}-f(\tau)\right) \left(\left(1-\hat{x}\right)-\gamma \hat{y}\right) \left( 
                1 - \hat{x} \right) - \iota, \hat{x}(0)= \hat{x}_0, \label{eqm1} \\
                \dot{\hat{y}} &= \kappa \left( \left(1-\hat{x}\right)^2 - \hat{y}^2 \right), \hat{y}(0)= \hat{y}_0, \label{eqm2} 
            \end{align}
            where $\gamma = 1.6,$ $\kappa = 0.1,$ $\iota = 0.009$, $\hat{x}_0=1-1$ x 10$^{-4}$, $\hat{y}_0 = 1$ x 10$^{-4}$, and $f(\tau)=\dfrac{1}{10}\left(\text{sin}\left(\dfrac{2 \pi \tau}{300}\right) + 1\right).$

Using this model, we can qualitatively match the body temperature recordings from the noisy miner \cite{miner}.





\subsubsection{The elephant shrew model}


We convert the noisy miner model (described by Eqs. (\ref{eqm1}) -(\ref{eqm2}) ) into a stochastic ODE model of the form

                \begin{align}
                \dot{\hat{x}} &= -\left(\hat{x}-f(\tau)\right) \left(\left(1-\hat{x}\right)-\gamma \hat{y}\right) \left( 
                1 - \hat{x} \right) - \iota \eta, \hat{x}(0)= \hat{x}_0, \label{eq3} \\
                \dot{\hat{y}} &= \kappa \left( \left(1-\hat{x}\right)^2 - \hat{y}^2 \right), \hat{y}(0)= \hat{y}_0, \label{eq4} 
            \end{align}

            where $\gamma = 1.6,$ $\kappa = 0.3,$ $\iota = 0.02$, $\hat{x}_0=1-1$ x 10$^{-4}$, $\hat{y}_0 = 1$ x 10$^{-4}$, $f(\tau)=\dfrac{1}{10}\left(\text{sin}\left(\dfrac{2 \pi \tau}{300}\right) + 1\right)$, and $\eta$ is a brownian noise process. Using this model, we can qualitatively match the body temperature recordings from an elephant shrew \cite{shrew}.

\subsubsection{The black bear model}


We modify the noisy miner model (described by Eqs.  (\ref{eqm1}) -(\ref{eqm2})) to include two forcing functions, $g(\tau)$ and $h(\tau)$, as follows:

                \begin{align}
                \dot{\hat{x}} &= -\left(\hat{x}-h(\tau)\right) \left(\left(g(\tau)-\hat{x}\right)-\gamma \hat{y}\right) \left( 
                g(\tau) - \hat{x} \right) - \iota, \hat{x}(0)= \hat{x}_0, \label{eq7} \\
                \dot{\hat{y}} &= \kappa \left( \left(g(\tau)-\hat{x}\right)^2 - \hat{y}^2 \right), \hat{y}(0)= \hat{y}_0, \label{eq8} 
            \end{align}
            where $\gamma = 1.5,$ $\iota = 0.0015,$ $\kappa = 0.1$, $\hat{x}_0=1-1$ x 10$^{-4}$, $\hat{y}_0 = 1$ x 10$^{-4}$, $h(\tau)=\dfrac{3}{10}\left(\text{sin}\left(\dfrac{2 \pi \tau}{10000}\right) + 1\right)$, and $g(\tau)=\dfrac{2}{10}\left(\text{sin}\left(\dfrac{2 \pi \tau}{10000}\right) + 4\right).$ Using this model, we can qualitatively match the body temperature recordings from a black bear \cite{toien}.



\section{Acknowledgements}

\subsection{Acknowledgments}
We thank Mary Silber, Danny Abrams, and Bard Ermentrout for helpful discussions regarding bifurcation theory and dynamical systems, Roan Luikart for his help digitizing data sets, and Alasdair Hastewell for his careful read and suggestions on the manuscript. C.E.F. also thanks the organizers and participants of the 2023 University of Alaska Fairbanks Hibernation Science Workshop. C.E.F. thanks Matthew Gorby of Research Computing Services at Northwestern University for computational assistance. 

\subsection{Funding}
  C. E. F. is supported in part by The James S. McDonnell Foundation Postdoctoral Fellowship Award in Complex Systems (\url{https://doi.org/10.37717/2020-1591}) and by the NSF-Simons Center for Quantitative Biology at Northwestern University (NSF: 1764421 and Simons Foundation/SFARI 597491-RWC). N. M. M. was supported by the U.S. Department of Energy, Office of Science, Office of Advanced Scientific Computing Research, under Award Number DESC0024253. A.J.E. was supported as a Quantitative Biology REU student at Northwestern University (NSF: 2150134). This research was supported in part by grants from the NSF (DMS-2235451) and Simons Foundation (MP-TMPS-00005320) to the NSF-Simons National Institute for Theory and Mathematics in Biology (NITMB). This research was supported in part through the computational resources and staff contributions provided for the Quest high performance computing facility at Northwestern University which is jointly supported by the Office of the Provost, the Office for Research, and Northwestern University Information Technology.

\section{Author contributions statement}

C.E.F. and N.M.M. designed research; C.E.F. and A.J.M. performed research; C.E.F and A.J.M. analyzed data; C.E.F. wrote the paper; A.J.M. and N.M.M. edited the paper.








\bibliography{main}

\end{document}